%% file: paper.tex
\documentclass[acmsmall,nonacm]{acmart}
\acmJournal{PACMPL}
\acmArticle{757}
\acmYear{2021}
\acmMonth{8}
\acmDOI{} 
\startPage{1}

\usepackage{booktabs}   
\usepackage{subcaption} 

\usepackage{xspace}
\usepackage{amsthm}
\usepackage[linesnumbered]{algorithm2e}
\usepackage{empheq}
\usepackage{amsmath}
\usepackage{pifont}
\usepackage{array}

\newcommand{\rlibm}{\textsc{RLibm}\xspace}
\newcommand{\Rlibmtt}{\textsc{RLibm-32}\xspace}
\newcommand{\tool}{\textsc{RLibm-All}\xspace}
\newcommand{\ourlibm}{\textsc{RLibm-All}\xspace}

\newcommand{\eg}{\emph{e.g.}\xspace}
\newcommand{\ie}{\emph{i.e.}\xspace}

\newcommand{\cmark}{\ding{51}}
\newcommand{\xmark}{\textcolor{red}{\ding{55}}}
\newcommand{\NA}{N/A}
\newcolumntype{P}[1]{>{\centering\arraybackslash}p{#1}}

\newcommand{\R}{\ensuremath{\ensuremath{\mathbb{R}}}\xspace}
\newcommand{\T}{\ensuremath{\mathbb{T}}\xspace}
\newcommand{\Tn}{\ensuremath{\mathbb{T}_{n}}\xspace}
\newcommand{\Tsmall}{\ensuremath{\mathbb{T}_{k}}\xspace}
\newcommand{\Tlarge}{\ensuremath{\mathbb{T}_{n+2}}\xspace}

\newcommand{\FP}{\ensuremath{\mathbb{F}_{n, |E|}}\xspace}
\newcommand{\FPK}{\ensuremath{\mathbb{F}_{k, |E|}}\xspace}

\newcommand{\FPInfty}{\ensuremath{\mathbb{F}_{\infty, |E|}}\xspace}

\newcommand{\Real}{\ensuremath{v_{\R}}\xspace}
\newcommand{\Prec}{\ensuremath{v^{-}}\xspace}
\newcommand{\Succ}{\ensuremath{v^{+}}\xspace}
\newcommand{\Rbit}{\ensuremath{rb}\xspace}
\newcommand{\Sticky}{\ensuremath{sticky}\xspace}

\newcommand{\Vrno}{\ensuremath{v_{rno}}\xspace}

\newcommand{\svrno}{$s_{v_{rno}}$\xspace}
\newcommand{\vmvrno}{$v^{-}_{v_{rno}}$\xspace}
\newcommand{\rbvrno}{$rb_{v_{rno}}$\xspace}
\newcommand{\stickyvrno}{$sticky_{v_{rno}}$\xspace}

\newcommand{\Vsm}{\ensuremath{v_{sm}}\xspace}
\newcommand{\Vlg}{\ensuremath{v_{lg}}\xspace}

\newcommand{\RNE}{\ensuremath{rn}\xspace}
\newcommand{\RNA}{\ensuremath{ra}\xspace}
\newcommand{\RNZ}{\ensuremath{rz}\xspace}
\newcommand{\RNP}{\ensuremath{ru}\xspace}
\newcommand{\RNN}{\ensuremath{rd}\xspace}
\newcommand{\RNO}{\ensuremath{ro}\xspace}

\newcommand{\Round}[3]{\ensuremath{RN_{#1, #2}(#3)}\xspace}

\newtheorem{theorem}{Theorem}

\newtheorem{lemma}{Lemma}

\theoremstyle{definition}

\begin{document}

\title[\tool: A Novel Polynomial Approximation Method to Produce Correctly Rounded Results for Multiple Representations and Rounding Modes]{\tool: A Novel Polynomial Approximation Method to Produce Correctly Rounded Results for Multiple Representations and Rounding Modes \\ {\small \textbf{Rutgers Department of Computer Science Technical Report DCS-TR-757}}}

\author{Jay P. Lim}
\orcid{0000-0002-7572-4017}             
\affiliation{
  \department{Department of Computer Science}
  \institution{Rutgers University}
  \country{United States}
}
\email{jpl169@cs.rutgers.edu}

\author{Santosh Nagarakatte}
\orcid{0000-0002-5048-8548}             
\affiliation{
  \department{Department of Computer Science}
  \institution{Rutgers University}
  \country{United States}
}
\email{santosh.nagarakatte@cs.rutgers.edu}

\input{sec.abstract.tex}

\maketitle
\input{sec.intro.tex}

\input{sec.background.tex}

\input{sec.illustration.tex}
\input{sec.new-approach.tex}

\input{sec.new-proof.tex}

\input{sec.eval.tex}
\input{sec.related.tex}

\input{sec.conclusion.tex}

\begin{acks}                            
  We thank John Gustafson for his inputs on the Minefield method and
  the posit representation.  This material is based upon work
  supported in part by the \grantsponsor{GS100000001}{National Science
    Foundation}{http://dx.doi.org/10.13039/100000001} under Grant
  No.~\grantnum{GS100000001}{1908798}, Grant
  No.~\grantnum{GS100000001}{1917897}, and Grant
  No.~\grantnum{GS100000001}{2110861}.
  Any opinions, findings, and
  conclusions or recommendations expressed in this material are those
  of the authors and do not necessarily reflect the views of the
  National Science Foundation.
\end{acks}

\bibliography{reference}

\end{document}

%% file: sec.abstract.tex
\begin{abstract}
Mainstream math libraries for floating point (FP) do not produce
correctly rounded results for all inputs. In contrast, CR-LIBM and
\rlibm provide correctly rounded implementations for a specific FP
representation with one rounding mode. Using such libraries for a
representation with a new rounding mode or with different precision
will result in wrong results due to double rounding.  This paper
proposes a novel method to generate a single polynomial approximation
that produces correctly rounded results for all inputs for multiple
rounding modes and multiple precision configurations. To generate a
correctly rounded library for $n$-bits, our key idea is to generate a
polynomial approximation for a representation with $n+2$-bits using
the \emph{round-to-odd} mode.  We prove that the resulting polynomial
approximation will produce correctly rounded results for all five
rounding modes in the standard and for multiple representations with
$k$-bits such that $|E| +1 < k \leq n$, where $|E|$ is the number of
exponent bits in the representation. Similar to our prior work in the
\rlibm project, we approximate the correctly rounded result when we
generate the library with $n+2$-bits using the round-to-odd mode. We
also generate polynomial approximations by structuring it as a linear
programming problem but propose enhancements to polynomial generation
to handle the round-to-odd mode.  Our prototype is the first 32-bit
float library that produces correctly rounded results with all
rounding modes in the IEEE standard for all inputs with a single
polynomial approximation. It also produces correctly rounded results
for any FP configuration ranging from 10-bits to 32-bits while also
being faster than mainstream libraries.
\end{abstract}

%% file: sec.intro.tex
\section{Introduction}
The floating point (FP) representation is widely used to approximate
real numbers. The two main attributes of the FP representation are its
dynamic range (\ie, the range of values that can be represented) and
precision (\ie, the accuracy of each value represented).  As some real
numbers cannot be accurately represented in the FP representation,
they need to be rounded to the nearest result according to the
rounding mode.  Further, FP performance is important in various
domains ranging from scientific computing to machine learning. Hence,
modern accelerators, processors, and systems have explored new
variants of the FP representation that vary the dynamic range and the
precision. Intel's Bfloat16~\cite{Kalamkar:bfloat:arxiv:2019} and
FlexPoint~\cite{Koster:flexpoint:NIPS:2017}, Nvidia's
TensorFloat32~\cite{nvidia:tensorfloat:online:2020}, Microsoft's
MSFP~\cite{Rouhani:msfp12:neurlips:2020}, Facebook's Log Number
System~\cite{Johnson:online:2018:facebook}, and
Posits~\cite{Gustafson:sfi:2017:beating} are examples of such recent
FP variants. All these representations need math libraries that
provide approximations for various elementary functions (\eg, $ln(x)$,
$e^x$, $\dots$)~\cite{Muller:elemfunc:book:2005}.

An output of an elementary function is a correctly rounded result if
it matches the result that is computed with infinite precision and
then rounded to the target FP representation. Correctly rounded
elementary functions improve the portability and reproducibility of
applications.
Unfortunately, mainstream FP libraries do not produce correct results
for all inputs.  Libraries like
CR-LIBM~\cite{Daramy:crlibm:spie:2003,Daramy:crlibm:doc} and
\rlibm~\cite{Lim:rlibm32:arxiv:2021, Lim:rlibm:arxiv:2020,
  lim:rlibm:popl:2021,lim:rlibm32:pldi:2021} provide correctly rounded
implementations for some FP representations. CR-LIBM provides
correctly rounded elementary functions for the \texttt{double} type
with a given rounding mode. As part of our \rlibm project, we have
developed correctly rounded libraries for the float, bfloat16, and
posit types with the round-to-nearest-ties-to-even mode.

Beyond the round-to-nearest-ties-to-even mode, there are four other
rounding modes in the IEEE-754 standard. When existing correctly
rounded libraries are used to generate results with other rounding
modes or with other precision configurations, they can produce wrong
results due to double rounding.  For example, let us say we use a
correctly rounded double precision library such as CR-LIBM with the
round-to-nearest-ties-to-even mode to produce results for a 32-bit
float type with the same rounding mode. Here, we round the result of
CR-LIBM to a 32-bit float value to produce the final result. If the
real value of $f(x)$ is extremely close to the rounding boundary of
two adjacent float values, then the error caused by rounding $f(x)$ to
double using the given rounding mode can be significant enough to
produce the wrong result for a 32-bit float.

With existing approaches such as \rlibm and CR-LIBM, one will have to
generate a new polynomial approximation for each such rounding mode
and each precision configuration. Although feasible, developing
efficient approximations require significant effort. Even after
decades of effort, there are no efficient and correctly rounded
implementations for all rounding modes in the IEEE standard even for
the widely used 32-bit float type!

\textbf{This paper.} Rather than generating a correctly rounded
elementary function for each individual representation and rounding
mode, it would be ideal if we could generate one polynomial
approximation that produces correct results for multiple rounding
modes and many precision configurations. This paper proposes a novel
approach to create such polynomial approximations! Our key idea is to
create polynomial approximations that approximate the correctly
rounded result of an elementary function $f(x)$ with the
\emph{round-to-odd} rounding mode (\ie, the real value of $f(x)$ is
rounded with the round-to-odd mode). The round-to-odd is a
non-standard rounding mode that can be described as follows.  If the
real value is exactly representable in the target representation, it
is unchanged. Otherwise, it is rounded to the nearest value in the
target representation whose bit-string when interpreted as an unsigned
integer is an odd number.

The round-to-odd mode has been previously used to address double
rounding issues in the context of binary FP to decimal FP
conversion~\cite{goldberg:fp} and also while performing primitive
arithmetic operations with extended precision~(\eg, Intel's 80-bit
floating point) and subsequently rounding the result back to a lower
precision (\ie, a float or a double type)~\cite{boldo:rno:imacs:2005,
  boldo:round-to-odd:tc:2008}.

This paper makes a case for using the round-to-odd mode to generate
correctly rounded results for elementary functions. To the best of our
knowledge, no prior method for approximating elementary functions has
used the round-to-odd mode.  Further, the usage of the round-to-odd
mode for approximating elementary functions is feasible because we
approximate the correctly rounded result in the \rlibm project. We
discover that the round-to-odd mode has properties that enable the
generation of correctly rounded elementary functions for multiple
rounding modes and multiple precision
configurations~(Section~\ref{sec:approach:rno}). 

If the goal is to produce correctly rounded elementary functions for
multiple rounding modes of a representation with $n$-bits~(\ie, \Tn),
we propose to generate polynomial approximations that produce the
correctly rounded results of the representation with
$n+2$-bits~(\Tlarge) using the round-to-odd mode.  We prove that this
polynomial approximation for $\Tlarge$ produces correctly rounded
results for all representations with $k$-bits (\ie, \Tsmall), where
$k$ is smaller than or equal to $n$ and for all rounding modes in the
IEEE standard. The only requirement in our proof is that the number of
exponent bits in both the representation used to create the
approximation (\ie, \Tlarge) and the target representation (\ie,
\Tsmall) is identical.
In summary, our approach that creates the math library with the
round-to-odd mode for a configuration with two additional bits
produces correctly rounded results for all target representations with
any standard rounding mode.

The next task is to generate polynomial approximations using the
round-to-odd mode for a representation with $n+2$-bits (\ie, \Tlarge).
We extend our prior work in the \rlibm project to generate polynomial
approximations with the round-to-odd mode.  Specifically, we
approximate the correctly rounded result of an elementary function
$f(x)$ with the round-to-odd mode rather than the real value of
$f(x)$.  Subsequently, we identify an interval of real values around
the correctly rounded result with the round-to-odd mode such that any
value in the interval rounds to the correctly rounded result, which we
call the odd interval. We show that any real value in the odd interval
that rounds to the round-to-odd result in \Tlarge will subsequently
round to the correctly rounded result for any representation \Tsmall.

One challenge in generating polynomial approximations using the odd
intervals is the presence of singleton values with the round-to-odd
mode. When the real value is exactly representable in \Tlarge and the
bit-string of that value in \Tlarge is even when interpreted as an
unsigned integer, the odd interval for that value will be a singleton
element. We use the mathematical properties of the elementary function
to identify such singletons. Concretely, we identify the inputs of an
elementary function that has rational outputs (because the
round-to-odd result is exactly representable in \Tlarge). We
subsequently develop efficient table-lookups to produce round-to-odd
results for such inputs~(Section~\ref{sec:approach:singleton}).

Once we identify the odd interval for every input and the singleton
odd intervals among them, we can subsequently use the \rlibm approach
to generate polynomial approximations.  Each non-singleton interval
imposes a constraint on the output of the generated polynomial for a
given input.
Similar to our prior work in the \rlibm
project~\cite{lim:rlibm:phdthesis:2021}, we structure the problem of
generating a polynomial approximation that satisfies the odd interval
for each input as a linear programming problem. We use efficient range
reduction and output compensation functions while accounting for
numerical errors with them. To account for numerical errors, we
further constrain the odd intervals given to the LP formulation. We
also employ counterexample guided polynomial generation and generate
piecewise polynomials for efficiency.

Our prototype, \ourlibm, is open-source and publicly
available~\cite{rlibm-all}. It contains the polynomial generator for
ten elementary functions and includes efficient polynomial
approximations for them.  The resulting polynomial generated by our
prototype for the FP representation with 34-bits (\ie,
$\mathbb{T}_{34}$) using the round-to-odd mode produces correctly
rounded results for all FP representations ranging from 10-bits (\ie,
$\mathbb{T}_{10}$) to 32-bits (\ie, $\mathbb{T}_{32})$ and for all
rounding modes. It includes bfloat16 and tensorfloat32
representations.  \ourlibm is the first math library that produces
correctly rounded results for 32-bit floats for all rounding modes
with a single polynomial approximation.
Our implementations are faster than mainstream libraries for 32-bit
floats while producing correctly rounded results for all inputs and
for all rounding modes.

%% file: sec.background.tex
\section{Background}
We provide background on the FP representation, the process of
rounding, and the various rounding modes in the IEEE-754 standard. As
we extend our \rlibm approach, we also provide a brief background on
generating polynomial approximations with it.

\input{fig.fp_rep.tex}

\subsection{The Floating Point Representation}
The floating point representation, which is specified by the IEEE-754
standard, is parameterized by the total number of bits~$n$ and the
number of bits for the exponent $|E|$, which we represent as
$\mathbb{F}_{n, |E|}$. The total number of bits and the number of bits
for the exponent determine the dynamic range and precision of the
representation. The FP bit-string consists of a sign bit, $|E|$ bits to
represent the exponent, and $n-1-|E|$ bits to represent the mantissa
($F$). Figure~\ref{fig:fp_rep} shows the bit-string for a standard
32-bit float and the custom 5-bit and 4-bit FP representations. If the
sign bit is $0$, then the value is positive. Otherwise, it is
negative. The value represented by the FP bit-string can be classified
into three classes: normal values, denormal values, and special
values.

The value represented by the FP bit-string is a normal value if the
bit-string $E$ is neither all ones nor all zeros (\ie, $0 < E <
2^{|E|} - 1$).  The normal value represented by this bit-string is $(1
+ \frac{F}{2^{|F|}})\times 2^{E-bias}$, where bias is $2^{|E| - 1} -
1$. If the exponent field $E$ is all zeros (\ie, $E = 0$), then the FP
value is a denormal value. Denormal values are used to represent
values close to zero. The real number represented by this denormal
value is $(\frac{F}{2^{|F|}})\times 2^{1-bias}$.
When the exponent field is all ones (\ie, $E=2^{|E|} - 1$), the FP
bit-strings represent special values. If $F = 0$, then the bit-string
represents $\pm \infty$ depending on the sign and in all other cases,
it represents \textit{not-a-number} (NaN).

The default FP types are the 16-bit half type ($\mathbb{F}_{16, 5}$),
the 32-bit float type ($\mathbb{F}_{32, 8}$), and the 64-bit double
type ($\mathbb{F}_{64, 11}$). Beyond these types, recent extensions
have increased the dynamic range and/or precision especially in the
context of machine learning.  The new types include
bfloat16~($\mathbb{F}_{16, 8}$)~\cite{Tagliavini:bfloat:date:2018} and
tensorfloat32~($\mathbb{F}_{19,
  8}$)~\cite{nvidia:tensorfloat:online:2020}.

\subsection{Rounding a Real Number to the FP Representation}
\label{sec:background:rounding}
\input{fig.fp_rounding_modes.tex}

Any FP representation can represent a finite number of real
values. Hence, many real values ($\Real$) cannot be exactly
represented. It is rounded to either the largest FP value smaller than
$\Real$ ($\Vsm$) or the smallest FP value larger than $\Real$
($\Vlg$).

\begin{align}
\small
\Vsm = max\{v \in \FP \mid v \leq \Real\} \nonumber \\
\Vlg = min\{v \in \FP \mid v \geq \Real\} \nonumber
\end{align}

The rounding mode, which we represent by $rm$, specifies whether
$\Real$ rounds to $\Vsm$ or $\Vlg$. We denote the operation of
rounding $\Real$ using a rounding mode $rm$ to a value in the
representation \FP with $\Round{\FP}{rm}{\Real}$. The IEEE-754
standard specifies five different rounding modes:
round-to-nearest-ties-to-even~(\RNE),
round-to-nearest-ties-to-away~(\RNA), round-towards-zero~(\RNZ),
round-towards-positive-infinity~(\RNP), and
round-towards-negative-infinity (\RNN). The standard mandates correct
rounding for primitives operations~(\ie, $+$, $-$, $*$, $/$).

\textbf{The round-to-nearest-ties-to-even (\RNE) mode.} This rounding
mode rounds \Real to a FP value that is closer to \Real among $\Vsm$
and $\Vlg$.  If $|\Vsm - \Real| < |\Vlg - \Real|$, then $\Real$ rounds
to $\Vsm$. If $|\Vsm - \Real| > |\Vlg - \Real|$, then $\Real$ rounds
to $\Vlg$. If \Real is exactly in the middle of $\Vsm$ and $\Vlg$
(\ie, $\Real = \frac{\Vsm + \Vlg}{2}$), $\Real$ is rounded to a value
whose bit-string is even when interpreted as an unsigned integer. The
\RNE mode is the most commonly used rounding
mode. Figure~\ref{fig:chap6:fp_rounding_modes}(a) and (b) illustrate
rounding with the \RNE mode depending on whether \Vsm is even or odd,
respectively.

\textbf{The round-to-nearest-ties-to-away (\RNA) mode.} This rounding
mode also rounds \Real to a FP value that is closer to \Real among
$\Vsm$ and $\Vlg$. If $|\Vsm - \Real| < |\Vlg - \Real|$, then $\Real$
rounds to $\Vsm$. If $|\Vsm - \Real| > |\Vlg - \Real|$, then $\Real$
rounds to $\Vlg$. When $\Real$ is exactly in the middle of $\Vsm$ and
$\Vlg$, $\Real$ is rounded to a value that is farther away from
$0$. Specifically, $\Real$ rounds to $\Vlg$ if $\Real > 0$ because $0
\leq \Vsm < \Real < \Vlg \leq \infty$.  Similarly, $\Real$ rounds to
$\Vsm$ if $\Real < 0$ because $ -\infty \leq \Vsm < \Real < \Vlg \leq
0$. Figure~\ref{fig:chap6:fp_rounding_modes}(c) and (d) illustrate
rounding with \RNA mode depending on whether $\Real < 0$ or $\Real >
0$, respectively.

\textbf{The round-towards-zero (\RNZ) mode.} In this mode, \Real is
rounded to a value that is closer to $0$. Here, \Real is rounded to
$\Vsm$ if $\Real > 0$ and \Real is rounded to $\Vlg$ if $\Real <
0$. The \RNZ mode is equivalent to truncating the fraction bits of
\Real that cannot fit within the mantissa bits of the
representation. Figure~\ref{fig:chap6:fp_rounding_modes}(e) and (f)
illustrate the \RNZ mode depending on whether $\Real < 0$ or $\Real >
0$, respectively.

\textbf{The round-towards-positive-infinity (\RNP) mode.} This mode
always rounds \Real to the larger value $\Vlg$, which is the value
that is closer to $+\infty$. This mode is also known as rounding up.
Figure~\ref{fig:chap6:fp_rounding_modes}(g) pictorially shows the \RNP
mode.

\textbf{The round-towards-negative-infinity (\RNN) mode.} The
round-towards-negative-infinity (\RNN) mode always rounds \Real to the
smaller value $\Vsm$ (\ie, a value that is closer to $-\infty$). This
mode is also known as rounding
down. Figure~\ref{fig:chap6:fp_rounding_modes}(h) demonstrates the
\RNN mode.

\subsection{A Systematic Method for Rounding}
We describe a systematic procedure for rounding a real number, which
will be useful later for understanding our proofs.  As we described
above, we need to identify the two values $\Vsm$ and $\Vlg$ in the
$\FP$ representation that are adjacent to \Real and then decide
between \Vsm or \Vlg. We will identify four pieces of information
($s$, \Prec, \Rbit, $\Sticky$) from the real value \Real that will be
sufficient to correctly round according to the various rounding
modes. We call them rounding components. The first component, $s$
represents the sign (-1 or 1) and identifies whether \Real is positive
or negative. The smaller of \Vsm or \Vlg in magnitude is represented
by \Prec.  The components $rb$ and $sticky$ encode information about
whether \Real is in the middle, closer to \Vsm, or closer to \Vlg.

To identify the rounding components, we represent \Real in the FP
representation with an infinite number of mantissa bits while having
the same number of exponents bits as \FP. We call this representation
extended infinite precision representation (\ie,
\FPInfty). Effectively, this extended precision representation is
similar to the \FP representation but has a large number of bits for
the mantissa. When \Real is larger than the dynamic range of \FP, we
represent it with the largest representable value in \FPInfty (\ie,
the exponents bits correspond to the largest normal value in \FP and
all the mantissa bits are ones). We cannot use $\infty$ to represent
$|\Real|$ because we need to make a clear distinction between a real
number and $\infty$. The round-towards-zero mode never rounds a real
value to $\infty$. Similarly, the round-towards-positive-infinity mode
does not round negative real values to $-\infty$ and the
round-towards-negative-infinity mode does not round positive real
values to $\infty$.

\begin{align}
\small
B_{\Real} = b_1 b_2 b_3 \dots b_{n} b_{n+1} b_{n+2} \dots 
\end{align}

Here, $b_1$ is the sign bit and bits $b_2 \dots b_{|E|+1}$ represent
the exponent bits. The rest of the bits starting from $b_{|E| + 2}
\dots$ represent the mantissa.

\input{fig.fp_prec_succ.tex}

\textbf{Identify rounding components.} To identify \Vsm and \Vlg that
\Real can round to, we identify two positive values $\Prec$ and
$\Succ$ adjacent to $|\Real|$ (\ie, the magnitude of \Real) in
\FP. Here, $\Prec$ represents the largest value that is smaller than
or equal to $|\Real|$ and $\Succ$ represents the smallest value larger
than $|\Real|$. To identify $\Prec$, we truncate $B_{|\Real|}$ to $n$
bits,
\[
\small
B_{\Prec} = 0 b_2 b_3 b_4 \dots b_{n-1} b_n
\]

Note that the sign bit is 0 because we are just considering the
magnitude for \Prec. We call \Prec the truncated value, which is a
rounding component. Then, the succeeding value of \Prec in \FP is
\Succ, which is obtained by adding 1 to \Prec. We maintain the
invariant: $\Prec \leq |\Real| < \Succ$. In the context of rounding
$\Real$ to \FP, $\Prec$ and $\Succ$ satisfy the following property,

\[
\small
\begin{cases}
 -\Succ < \Real \leq -\Prec & \text{if } \Real < 0 \text{ } (s = -1) \\
\Prec \leq \Real < \Succ & \text{if } \Real \geq 0 \text{ } (s = 1)
\end{cases}
\]

Once we identify $s$, \Prec, and \Succ, we can compute \Vsm and \Vlg
as follows. If \Real is exactly representable in \FP, then $|\Real| =
\Prec$. Hence, $\Vsm = \Vlg = s \times \Prec$. If \Real is not exactly
representable in \FP, then it is guaranteed that $\Prec < |\Real| <
\Succ$. Thus, $\Vsm = \Prec$ and $\Vlg = \Succ$ if $s = 1$ (\ie,
$\Real \geq 0$). Otherwise, when \Real is negative (\ie, $s = -1$)
then $\Vsm = -\Succ$ and $\Vlg = -\Prec$.

\vspace{4pt} \textbf{Rounding bit.} To determine the rounding decision
for the \RNE and \RNA mode, we must determine whether $\Real$ is
closer to $s \times \Prec$, closer to $s \times \Succ$, or exactly in
the middle of the two values. We extract the $(n+1)^{th}$-bit from our
extended precision representation of $B_{|\Real|}$, which we call as
the rounding bit (\ie, \Rbit).  The rounding bit describes whether
$|\Real|$ is closer to $\Prec$ than $\Succ$. If the rounding bit is
$0$, then $|\Real|$ is closer to $\Prec$ (\ie, $\Prec \leq |\Real| <
\frac{\Prec + \Succ}{2}$). If the rounding bit is $1$, then $|\Real|$
is at the middle or close to
$\Succ$. Figure~\ref{fig:chap6:fp_prec_succ} illustrates the range of
real values where the rounding bit is 0 or 1.

\vspace{4pt} \textbf{Sticky bit.} While the rounding bit tells us
whether $|\Real|$ is closer to \Prec, it does not tell us whether
$|\Real|$ is exactly equal to \Prec or is exactly in the middle of
\Prec and \Succ (\ie, $|\Real| = \frac{\Prec + \Succ}{2}$).  When we
look at the bit-string $B_{|\Real|}$, $|\Real|$ is equal to \Prec when
the $(n+1)^{th}$-bit (\ie, \Rbit) is 0 and the remaining bits from the
$(n+2)^{th}$-bit are all zeros. If $\Rbit = 0$ and any bit afterwards
is 1, then $|\Real|$ is not equal to \Prec. Similarly, $|\Real|$ is
exactly in the middle of \Prec and \Succ when the $(n+1)^{th}$-bit is
1 and the remaining bits from the $(n+2)^{th}$-bit are all 0's in
$B_{|\Real|}$. In both these cases, we need to determine if all the
bits starting from $b_{n+2}$ are zeros. We define the sticky
bit as the bitwise \texttt{OR} of all bits starting from the
$(n+2)^{th}$-bit in the extended precision representation.
\[
\small \Sticky =  b_{n+2} \:|\: b_{n+3} \:|\: b_{n+3} \:|\: \dots
\]
where $|$ is the bit-wise \texttt{OR} operation.

Using the rounding components $(s, \Prec, \Rbit, \Sticky)$, we can
identify the relationship between $|\Real|$ and the nearest FP values
for any rounding mode in the standard.

\[
\small
\begin{cases}
|\Real| = \Prec & \text{if } \Rbit = 0 \land \Sticky = 0 \\
\Prec < |\Real| < \frac{\Prec + \Succ}{2} & \text{if } \Rbit = 0 \land \Sticky = 1 \\
|\Real| = \frac{\Prec + \Succ}{2} & \text{if } \Rbit = 1 \land \Sticky = 0 \\
\frac{\Prec + \Succ}{2} < |\Real| < \Succ & \text{if } \Rbit = 1 \land \Sticky = 1
\end{cases}
\]

We can compute \Succ from the rounding
components. Figure~\ref{fig:chap6:fp_prec_succ} pictorially shows the
rounding bit and sticky bit for various real values between \Prec and
\Succ.

\input{fig.rne_rounding_decision.tex}

\input{fig.rm_rounding_decision.tex}

\textbf{Rounding to various modes with the rounding components.}
Figure~\ref{fig:chap6:fp_rne_rounding_decision} shows rounding \Real
using the rounding components for the \RNE mode. Similarly,
Figure~\ref{fig:chap6:fp_rm_rounding_decision} illustrates rounding
with the rounding components for the other four rounding modes.

\subsection{The \rlibm Approach}

We provide a brief background on our \rlibm project as we build on top
of it in this paper. In the \rlibm project, we make a case for
approximating the correctly rounded result rather than the real value
of an elementary function~\cite{lim:rlibm32:pldi:2021,
  Lim:rlibm:arxiv:2020,lim:rlibm:popl:2021,Lim:rlibm32:arxiv:2021}.
When we approximate the correctly rounded result, there is an interval
of real values around the correctly rounded result for each input such
that producing any value in the interval produces the correct
result. This interval is further constrained to account for numerical
errors that can occur with polynomial evaluation, range reduction, and
output compensation. This interval can be used to generate polynomial
approximations. It represents the maximum amount of freedom available
to produce the correct result. Figure~\ref{fig:rlibm_explain}
illustrates our \rlibm approach.

The \rlibm approach consists of four steps. The first step is to use
an oracle to compute the correctly rounded result of an elementary
function $f(x)$ for each input $x \in \mathbb{T}$, where $\mathbb{T}$
is the target representation.
The second step is to identify an interval $[l, h]$ around the
correctly rounded result such that any value in $[l, h]$ rounds to the
correctly rounded result in $\mathbb{T}$, which is known as the
\textit{rounding interval}. Since polynomial evaluation, range
reduction, and output compensation happen in representation with
higher precision $\mathbb{H}$, the rounding intervals are also in
$\mathbb{H}$.
The third step is to employ range reduction to transform input $x$ to
$x'$. The generated polynomial will approximate the result for
$x'$. Subsequently, we use output compensation to produce the final
correctly rounded output for $x$.  Both range reduction and output
compensation happen in $\mathbb{H}$ and can experience numerical
errors. These numerical errors should not affect the generation of
correctly rounded results.  Hence, it is necessary to deduce intervals
for the reduced domain so that the polynomial evaluation over the
reduced input produces the correct results for the original inputs.
Given $x$ and its rounding interval $[l, h]$, reduced input $x'$ is
computed with range reduction. The next task before polynomial
generation is to identify the reduced rounding interval for $P(x')$
that when used with output compensation produces the correctly rounded
result. We use the inverse of the output compensation function to
identify the reduced interval $[l', h']$.

\input{fig.rlibm_explain.tex}

The last step is to synthesize a polynomial of a degree $d$ using an
arbitrary precision linear programming (LP) solver that satisfies the
constraints (\ie, $l' \leq P(x') \leq h'$) when given a set of inputs
$x'$. Approximating the correctly rounded result with the \rlibm
approach provides more freedom in generating polynomials. Hence, the
resulting \rlibm libraries are more efficient compared to mainstream
libraries.

%% file: fig.fp_rep.tex
\begin{figure}[t]%
	\includegraphics[width=0.95\textwidth]{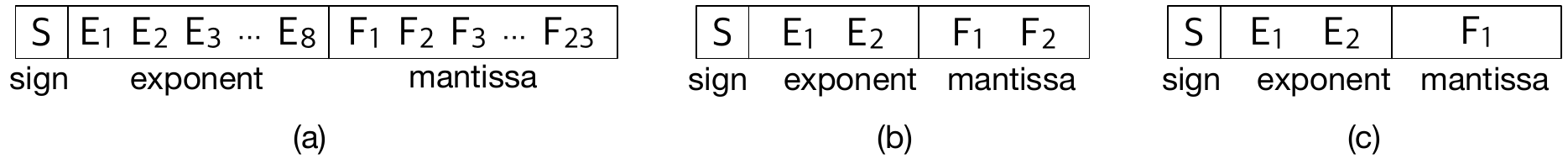}
	\caption{(a) The bit-string of a 32-bit float. (b) The
          bit-string of a 5-bit FP representation with 2 exponent and
          2 mantissa bits. (c) The bit-string of a 4-bit FP
          representation with 2 exponent and 1 mantissa bits.}
	\label{fig:fp_rep}
\end{figure}

%% file: fig.fp_rounding_modes.tex
\begin{figure}[t]%
	\includegraphics[width=\textwidth]{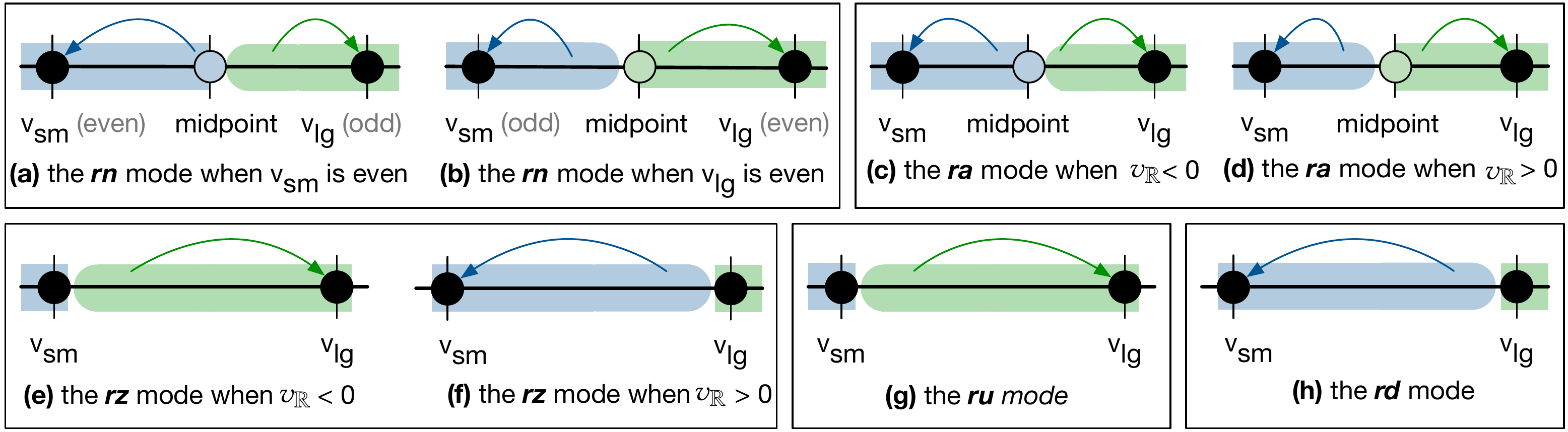}
	\caption{When a real value $\Real$ is not exactly
          representable in $\T$, then $\Real$ is rounded to one of the
          two adjacent values $\Vsm, \Vlg \in \T$ depending on the
          rounding mode. We show the range of real values (\Real) that
          round to $\Vsm$ (blue box) and $\Vlg$ (green box). (a) The
          \RNE mode when \Vsm is even. (b) The \RNE mode when \Vlg is
          even. (c) The \RNA mode when $\Real < 0$. (d) The \RNA mode
          when $\Real > 0$. (e) The \RNZ mode when $\Real < 0$. (f)
          The \RNZ mode when $x > 0$. (g) The \RNP mode. (h) The \RNN
          mode.}
	\label{fig:chap6:fp_rounding_modes}
\end{figure}

%% file: fig.fp_prec_succ.tex
\begin{figure}[t]%
	\includegraphics[width=\textwidth]{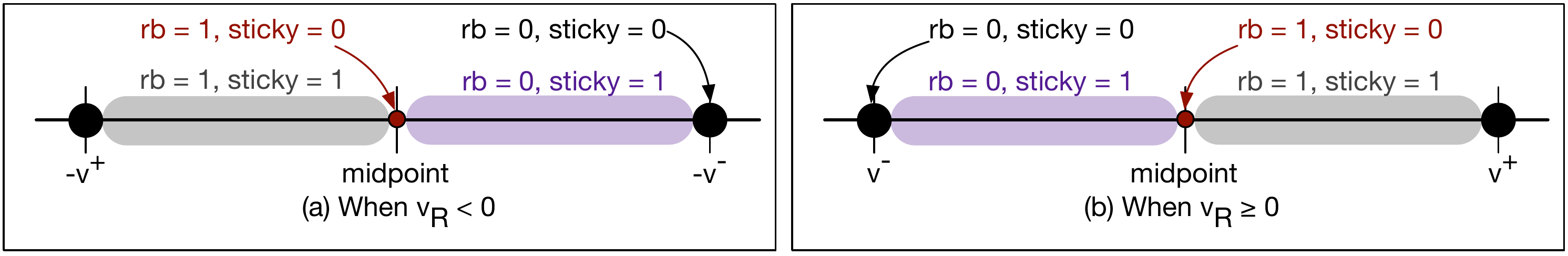}
	\caption{The values of the rounding bit and the sticky bit for
          various real values between \Prec and \Succ. The gray box
          indicates the range of real values where both \Rbit and
          \Sticky are 1. The purple box indicates the range of real
          values where \Rbit = 0 and \Sticky = 1.}
	\label{fig:chap6:fp_prec_succ}
\end{figure}

%% file: fig.rne_rounding_decision.tex
\begin{figure}[t]%
	\includegraphics[width=\textwidth]{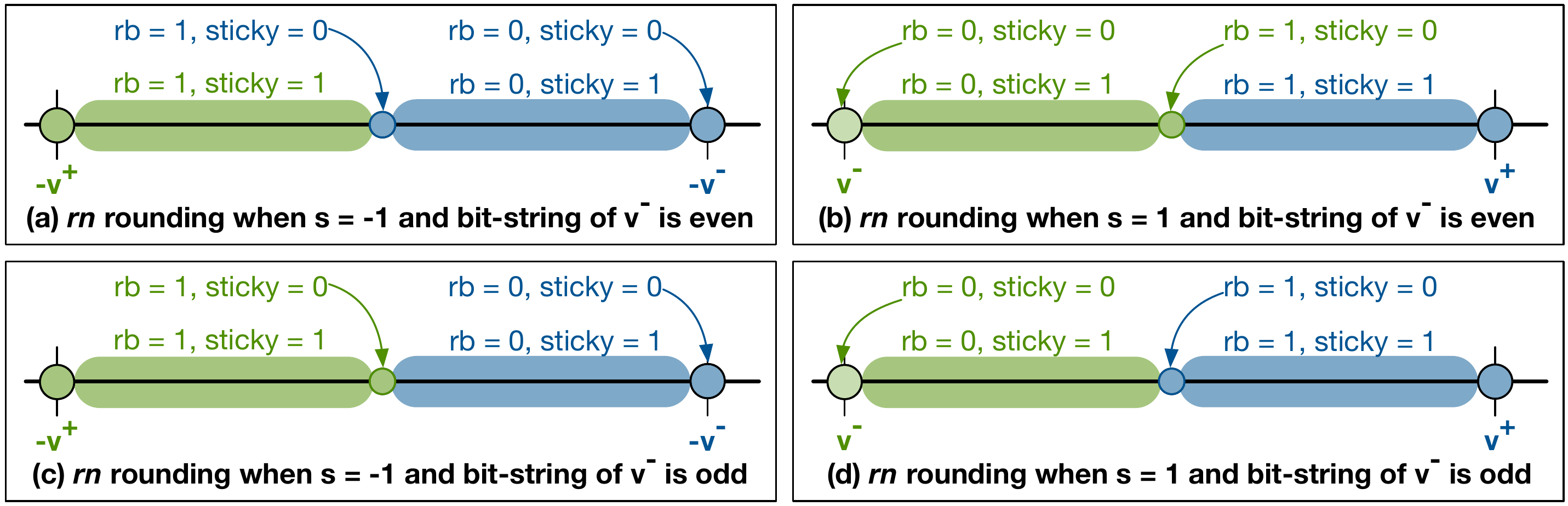}
	\caption{The \RNE mode using the rounding components ($s$,
          \Prec, \Rbit, \Sticky).  We illustrate rounding when \Real
          is positive or negative and when \Prec is even or odd. When
          a interval is colored green, all real values in the interval
          round to the FP value colored green. Similarly, all real
          values in the interval colored blue will round to the FP
          value colored blue.}
	\label{fig:chap6:fp_rne_rounding_decision}
\end{figure}

%% file: fig.rm_rounding_decision.tex
\begin{figure}[t]%
	\includegraphics[width=\textwidth]{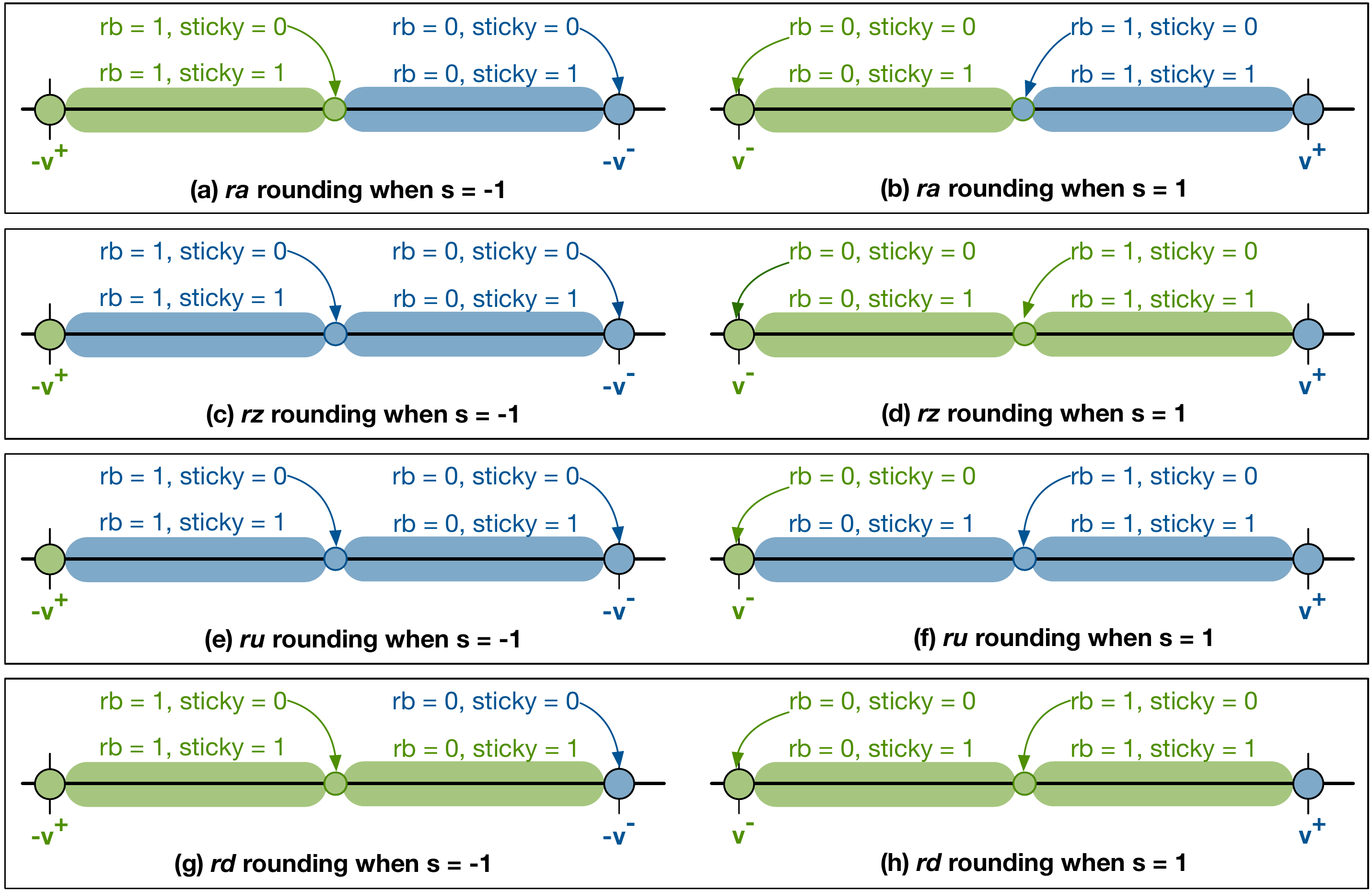}
	\caption{Rounding decisions for various rounding modes based
          on the rounding components: ($s$, \Prec, \Rbit, \Sticky).
          The interval of real values colored with green and blue
          round to the FP value colored green and blue, respectively.}
	\label{fig:chap6:fp_rm_rounding_decision}
\end{figure}

%% file: fig.rlibm_explain.tex
\begin{figure}[t]%
	\includegraphics[width=\textwidth]{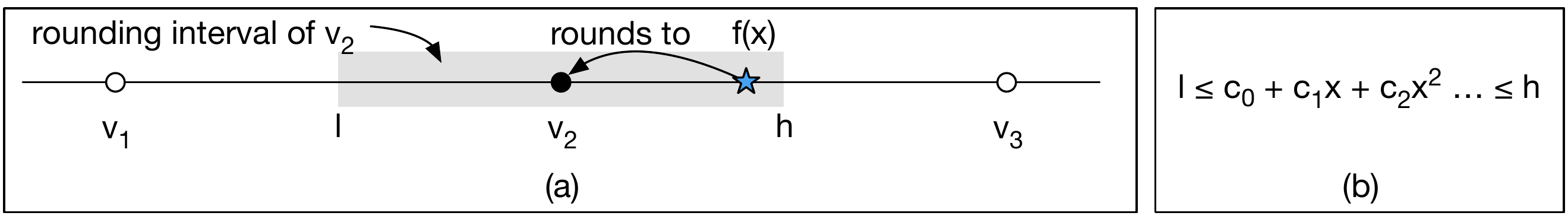}
	\caption{Illustration of the RLibm approach. (a) The values
          $v_1$, $v_2$, and $v_3$ are representable values in
          representation \T. The real value of $f(x)$ for a given
          input $x$ cannot be exactly represented in \T and it is
          rounded to $v_2$. The RLibm approach identifies the rounding
          interval of $v_2$ (shown in gray box). (b) Polynomial
          generation using the rounding interval (\ie, $[l, h]$) for
          each input $x$ with an LP formulation.}
	\label{fig:rlibm_explain}
\end{figure}

%% file: sec.illustration.tex
\section{An Illustrative Example of Our Approach}
\label{sec:illustration}
We describe our entire approach with an end-to-end example for
creating a polynomial approximation for $ln(x)$ that produces
correctly rounded results for a 5-bit FP representation with 2
exponent bits (FP5) and a 4-bit FP with 2 exponent bits (FP4) for all
standard rounding modes (\ie, \RNE, \RNA, \RNZ, \RNP, and
\RNN). Figure~\ref{fig:fp_rep}(b) and Figure~\ref{fig:fp_rep}(c) show
the bit-string of FP5 and FP4, respectively.  Although we illustrate
our approach with FP5 and FP4 for ease of exposition, it is beneficial
in practice to create table-lookups for FP5 and FP4 because there are
only 32 and 16 distinct bit-patterns, respectively.

The $ln(x)$ function is defined over the input domain $(0,
\infty)$. The result of $ln(x)$ is \texttt{NaN} when $x < 0$ or when x
is NaN. The result is $\infty$ when the input is $\infty$ and
$-\infty$ when the input is 0. There are only 11 non-special case
inputs, which range from $0.25$ to $3.5$ in FP5. Similarly, there are
only 5 non-special case inputs in FP4. Now, our goal is to generate a
single polynomial approximation that produces correctly rounded
results for both FP5 and FP4 with all five rounding modes.

To accomplish this goal, we will generate a polynomial approximation
that produces correctly rounded results for a 7-bit FP representation
(FP7) with the round-to-odd mode. Here, FP7 has exactly the same
number of exponent bits as FP5 and FP4 (\ie, 2 exponent
bits). Effectively, FP7 has 2 additional fraction bits when compared
to FP5. Every value that is representable in FP5 and FP4 is also
representable in FP7.  While rounding with the round-to-odd mode, if
the polynomial approximation produces a value that is exactly
representable in FP7, then it is unchanged. Otherwise, the result of
the polynomial approximation (which is implemented in double
precision) is rounded to the nearest FP7 value whose bit-string is odd
(\ie, the last bit is a 1). When this FP7 round-to-odd result is
rounded to a value in FP5 or FP4 according to any of the five standard
rounding modes, it produces the correct result for them.

\input{fig.rnd_int_ln1p5.tex}

\textbf{Why does a correctly rounded result with the round-to-odd mode
  for FP7 work with FP5/FP4?}  As the number of exponent bits is
identical in FP7, FP5, and FP4, every value that is representable in
FP5 and FP4 is also representable in FP7. Let us consider an input
$1.5$.  We want to produce correctly rounded results of $ln(1.5)$ for
all the rounding modes with FP5 and FP4.
The first row of Figure~\ref{fig:rnd_int_ln1p5} shows the real result
(\ie, a star) and the correctly rounded result of FP5 with the \RNE
mode. If we want to generate the correctly rounded result for
$ln(1.5)$ in FP5 with the \RNE mode using polynomial approximations,
there is an interval of real values around the correctly rounded
result such that producing any value in that interval produces the
correct result~(shaded in gray). The subsequent rows show the
correctly rounded result and the rounding intervals for other rounding
modes of FP5 and FP4 for the same input $1.5$. When compared to FP5,
the rounding interval for FP4 will be larger because the distance
between adjacent points is larger.  Intuitively, a single polynomial
can produce the correctly rounded result of $ln(1.5)$ for both FP5 and
FP4 with all rounding modes if it produces a value that lies in the
common interval among all these modes and precision configurations
(\ie, an intersection of the rounding intervals).

We show that computing the correctly rounded result of $ln(1.5)$ with
FP7 using the round-to-odd mode and identifying the interval around
this result in FP7 is an effective way to compute the common interval
described above. The last row of Figure~\ref{fig:rnd_int_ln1p5} shows
the correctly rounded result in FP7 with the round-to-odd mode and the
interval to produce that value. The interval for the round-to-odd
result in FP7 is smaller than the common interval among FP5 and FP4
with all the rounding modes because it works for many other
representations beyond FP5 and FP4.

In FP7, there are three additional values between the two adjacent FP5
values. Hence, the round-to-odd result with FP7 preserves enough
information to produce the correctly rounded result with FP5 and FP4
with any rounding mode. Our proofs in Section~\ref{sec:rno:proof} show
that this is a generic result for any representation with $n$-bits.

\textbf{Generating polynomial approximations.} The first step is to
identify rounding intervals for producing the correctly rounded result
of FP7 with the round-to-odd mode, which we call the odd interval. In
our approach, polynomial evaluation happens with double
precision. Hence, we identify an interval of values in double
precision such that any value in that interval rounds to the correctly
rounded result in FP7 with the round-to-odd mode. When the correctly
rounded result in FP7 with the round-to-odd mode is even (\ie, the
bit-string is even when interpreted as an unsigned integer), the
rounding interval is a singleton. For example, the odd interval for
$ln(1.0)$ is a singleton because the correctly rounded result is 0. If
the correctly rounded result with the round-to-odd mode is not even,
then we can identify the odd interval as follows. We identify the
preceding value (\emph{l}) and the succeeding (\emph{h}) value
corresponding to the correctly rounded result in FP7.  Then, the open
interval $(l,h)$ is the odd interval.  Figure~\ref{fig:geninterval}(b)
shows the odd interval for each input (shaded in blue). Next, we need
to create a polynomial approximation that produces a value in the odd
interval for each input.

Creating polynomial approximations with singleton odd intervals is
challenging because there is no freedom for the polynomial
generator. For the $ln(x)$ function with FP7, there is only one input
(\ie, $1.0$) whose odd interval is a singleton.  We treat it as a
special case. In general, we use mathematical properties of the
elementary function for larger data types to effectively handle such
singleton odd intervals
(Section~\ref{sec:approach:singleton}). Figure~\ref{fig:geninterval}(b)
shows the remaining inputs and their non-singleton odd intervals.

\input{fig.geninterval.tex}

The next step is to generate a polynomial that produces a value in the
odd interval for all inputs. We show the constraints imposed by the
odd interval on the output of the polynomial in
Figure~\ref{fig:geninterval}(a). As there only 10 non-special case
inputs, we encode them as a system of linear inequalities similar to
our prior work in the \rlibm project and use an LP solver to solve for
the coefficients of a $4^{th}$ degree polynomial $P(x)$ (see
Figure~\ref{fig:geninterval}(a)).  For larger representations, we
employ sophisticated range reduction, counterexample guided polynomial
generation, and generate piecewise
polynomials~\cite{lim:rlibm32:pldi:2021}.
Figure~\ref{fig:geninterval}(b) pictorially shows the generated
polynomial, which produces a value in the odd interval for each input.
This polynomial will produce the correctly rounded result of $ln(x)$
when the result is rounded to FP5 and FP4 with any of the five
rounding modes in the standard.

%% file: fig.rnd_int_ln1p5.tex
\begin{figure}[!t]%
  \includegraphics[height=2in]{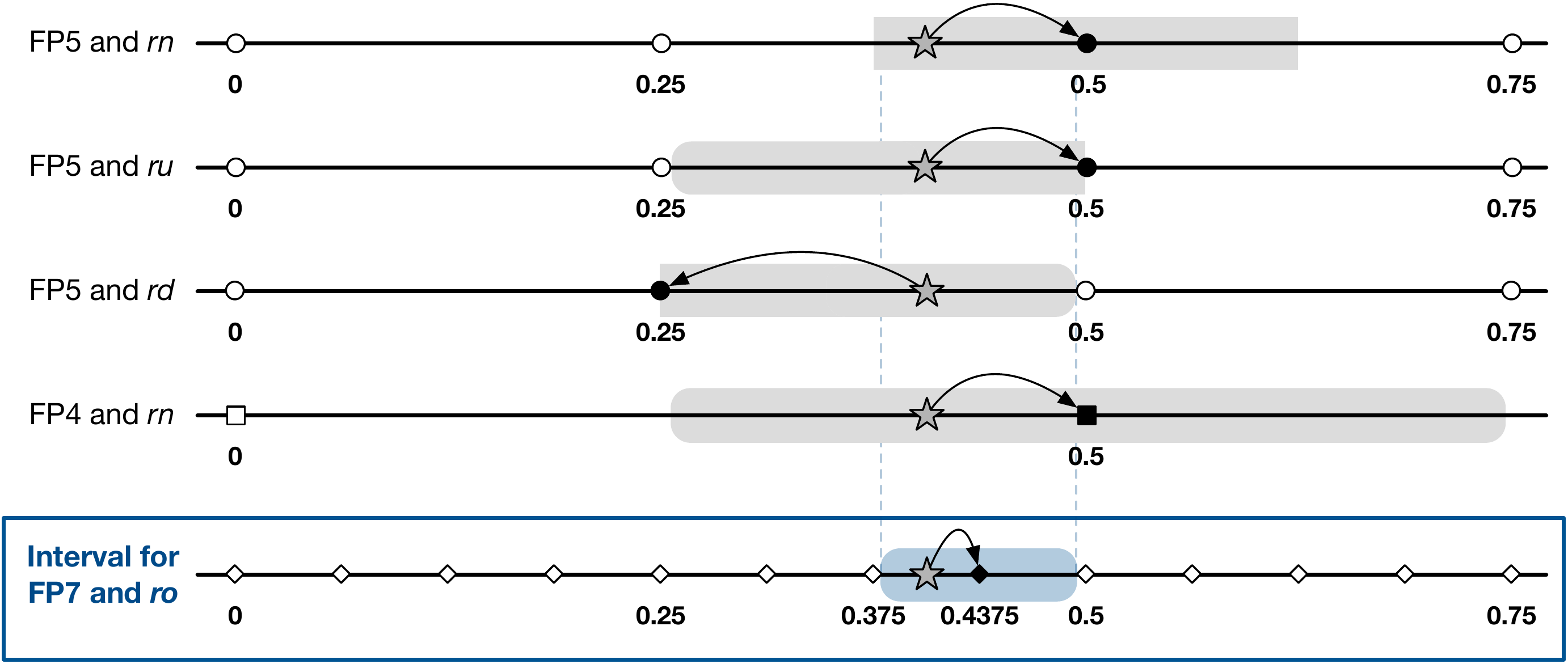}
  \caption{The correctly rounded result of $ln(1.5)$ for FP5 and FP4
    with some subset of the rounding modes and their rounding
    intervals (gray box). The gray star represents the real value of
    $ln(1.5)$. Values that are representable in FP7, FP5, and FP4 are
    shown with rhombus, circle, and square, respectively. Solid shapes
    represents the correctly rounded result for the chosen
    representation and rounding mode. The last row shows the odd
    interval to produce the correctly rounded result of $ln(1.5)$ in
    FP7 with the round-to-odd mode. The odd interval is a subset of the
    intersection of the rounding intervals of these configurations.}
  \label{fig:rnd_int_ln1p5}
\end{figure}

%% file: fig.geninterval.tex
\begin{figure}[!t]
	\begin{subfigure}[b]{0.45\linewidth}
		\centering
		\includegraphics[width=0.90\textwidth]{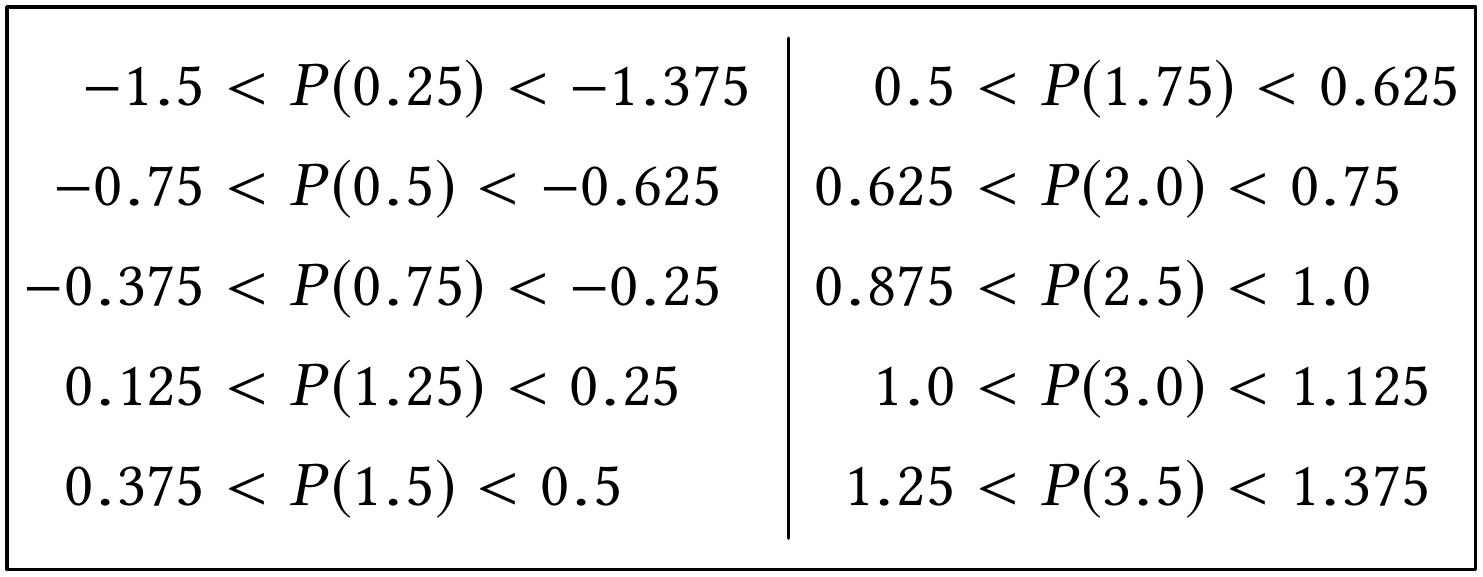}
		\vspace{1em}
		\caption{}
	\end{subfigure}
	\begin{subfigure}[b]{0.54\linewidth}
		\centering
		\includegraphics[width=\textwidth]{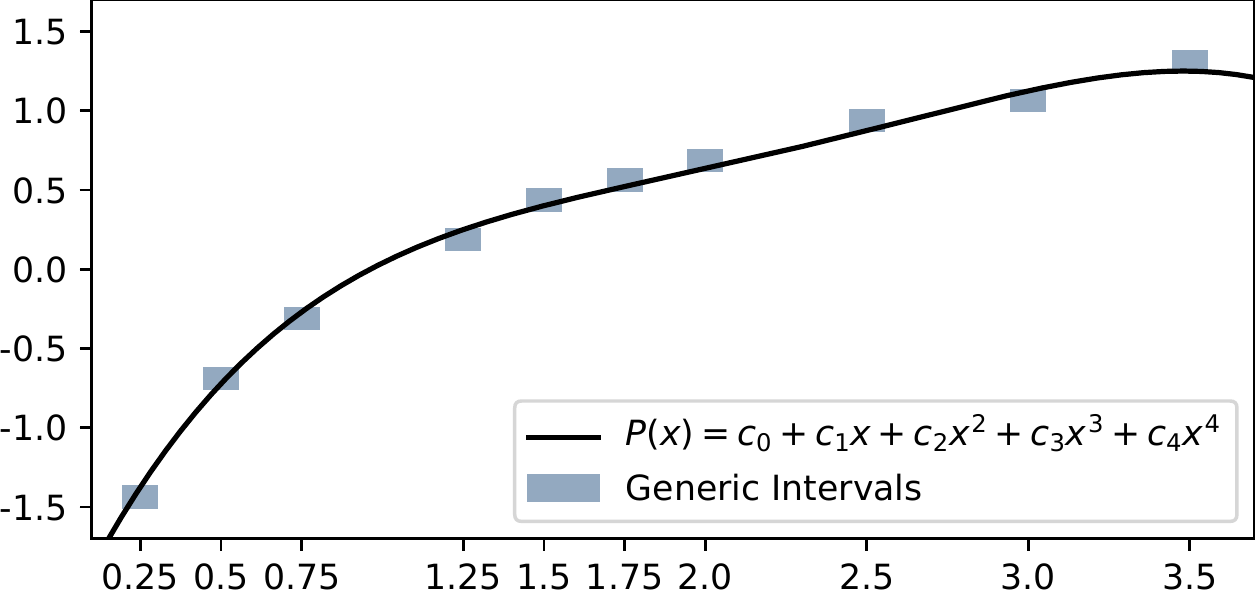}
		\caption{}
	\end{subfigure}
	\caption{(a) The set of constraints for the polynomial
          approximation to produce correctly rounded results for each
          input with the round-to-odd mode in FP7. (b) The odd
          intervals for each input and the resulting polynomial from
          our approach that produces a value in the odd interval for
          all inputs.}
  \label{fig:geninterval}
\end{figure}

%% file: sec.new-approach.tex
\section{Our Approach to Generate a Generic Polynomial Approximation}
\label{sec:approach}
Our goal is to generate a single polynomial approximation of an
elementary function that produces correctly rounded results for all
inputs for multiple precision and rounding configurations. Let $\Tn$
be a $n$-bit FP representation (\ie, \FP). Let $\Tsmall$ be a
representation where $\Tsmall$ has no more precision bits compared to
$\Tn$ with the same number of exponent bits. Specifically, $\Tsmall =
\FPK$ where $|E| + 1 < k \leq n$. Note that all values exactly
representable in $\Tsmall$ are also exactly representable in $\Tn$
(\ie, $\Tsmall \subseteq \Tn$). We define $rm$ to be a rounding mode
in the standard (\ie, $rm \in \{\RNE, \RNA, \RNZ, \RNP, \RNN\}$). Our
goal is to generate a polynomial approximation $A_{\mathbb{H}}(x)$,
which is implemented in representation $\mathbb{H}$, of an elementary
function $f(x)$ that produces correctly rounded results for all inputs
for any representation \Tsmall and any rounding mode
$rm$. Specifically, rounding the result of $A_{\mathbb{H}}(x)$ to any
representation $\Tsmall$ with the $rm$ rounding mode must result in
the same value as computing $f(x)$ in real numbers and rounding the
result to $\Tsmall$ with the $rm$ rounding mode, for all inputs in
$\Tsmall$.

\[
\small
\Round{\Tsmall}{rm}{A_{\mathbb{H}}(x)} = \Round{\Tsmall}{rm}{f(x)}
\]

\textbf{Main insight.} \emph{To generate correct results for \Tsmall,
our key insight is to create a polynomial approximation that produces
the correctly rounded results for $\Tlarge$ with the round-to-odd
mode}. We prove that it produces the correctly rounded result for any
representation $\Tsmall$ with all standard rounding modes when we
round the round-to-odd result to the target
representation~(Section~\ref{sec:rno:proof}). Intuitively, our
approach works because the round-to-odd result with $\Tlarge$
maintains sufficient information about the real value of an elementary
function $f(x)$ that is required for correct rounding for all
representations $\Tsmall$ with any standard rounding mode.

To generate a polynomial approximation for $\Tlarge$ with the
round-to-odd mode, we use our \rlibm approach. We approximate the
correctly rounded result rather than the real value. We extend the
\rlibm approach to handle the round-to-odd mode. Specifically, we need
to generate an interval of values around the correctly rounded
round-to-odd result for each input, which we call the odd interval. If
the generated polynomial produces a value in the odd interval for a
particular input, then it produces the correct result for all
representations $\Tsmall$ and for all rounding modes. One unique
challenge that we address is the presence of singleton odd intervals,
which happens when the correctly rounded result in $\Tlarge$ is
even. To scale to 32-bit floats (\ie, $\Tlarge$ = 34-bit float), we
employ range reduction, counterexample guided polynomial generation,
and generate piecewise polynomials. Finally, we use a linear
programming solver to solve for the coefficients of a polynomial given
a system of linear constraints generated from the odd intervals.

\subsection{Generating the Correctly Rounded Result for \Tlarge with the Round-to-Odd Mode}
\label{sec:approach:rno}
As we make a case for creating polynomial approximations for \Tlarge
with the round-to-odd mode, we formally define it and describe
rounding with the round-to-odd mode using the rounding components. We
describe the properties of the correctly rounded result with the
round-to-odd mode and provide intuition on why rounding the
round-to-odd result in \Tlarge to any representation \Tsmall produces
the correct result.

\input{fig.rno_intuition.tex}

The round-to-odd (\RNO) is a non-standard rounding mode that has been
previously used to avoid double rounding issues while converting a
binary FP number to a decimal FP number~\cite{goldberg:fp} and while
performing primitive operations~\cite{boldo:rno:imacs:2005,
  boldo:round-to-odd:tc:2008}. Given a real value \Real, the
round-to-odd mode rounds \Real as follows. If \Real is exactly
representable as value $v$ in the target representation, then \Real
rounds to $v$. Otherwise, \Real rounds to the nearest odd value in the
target representation.  Figure~\ref{fig:rno_intuition} illustrates the
round-to-odd rounding mode. Using the rounding components ($s$, \Prec,
\Rbit, \Sticky) from Section~\ref{sec:background:rounding}, the
round-to-odd mode can be defined as follows:
  
\[
\small
\Vrno = \Round{\mathbb{T}}{\RNO}{\Real} = 
\begin{cases}
s \times \Prec & \text{if } IsOdd(\Prec) \lor (\Rbit = 0 \land \Sticky = 0) \\
s \times \Succ & otherwise
\end{cases}
\]
where \Succ is the adjacent value to \Prec in $\mathbb{T}$.

Our contribution is to use the round-to-odd mode to generate correctly
rounded elementary functions for multiple representations and multiple
types. Specifically, we prove the following theorem in
Section~\ref{sec:rno:proof}, which forms the foundation for our
approach.

\begin{theorem}
\label{thm:main}
Let $\Real = f(x) $ be the real valued result of an elementary function
and $\Vrno = \Round{\Tlarge}{\RNO}{\Real}$. Let $v$ be a value in the
odd interval of $\Vrno$. Consider a rounding mode $rm \in$ \{\RNE,
\RNA, \RNZ, \RNP, \RNN\}. Then,
\[
\small
\Round{\Tsmall}{rm}{v} = \Round{\Tsmall}{rm}{\Real}
\]
\end{theorem}

We propose an efficient procedure to create polynomial approximations
$A_{\mathbb{H}}(x)$ of an elementary function $f(x)$ that produces
values in the odd interval of the correctly rounded result in \Tlarge.
Using Theorem~\ref{thm:main}, rounding any value $v$ in the odd
interval (\ie, $A_{\mathbb{H}}(x)$) to \Tsmall using a rounding mode
$rm$ produces the correctly rounded result of $f(x)$ in \Tsmall using
the same rounding mode $rm$.

\input{fig.rno_to_w1.tex}

\textbf{An example to show why the round-to-odd result avoids double
  rounding errors.} We provide intuition on how rounding with the
round-to-odd mode avoids double rounding errors in
Figure~\ref{fig:rno_to_w1}.  Any value that is representable in \Tn is
also representable in \Tlarge. Further, there are three additional
values ($w_1, w_2, w_3$) in \Tlarge between $w_0$ and $w_4$. Here,
$w_0$ and $w_4$ are also representable in \Tn.  In the round-to-odd
mode, any real value between $w_0$ and $w_2$ rounds to
$w_1$. Similarly, any real value between $w_2$ and $w_4$ rounds to
$w_3$. If the real value is exactly equal to $w_0$, then the
round-to-odd mode with \Tlarge also rounds to $w_0$ (similarly $w_2$
and $w_4$ with \Tlarge). Figure~\ref{fig:rno_to_w1} illustrates the
task of rounding the real value directly to \Tn with the \RNE mode
(solid arrow) and the result produced from double rounding the \RNO
result from \Tlarge to \Tn using the \RNE mode.

In the context of rounding a real value directly to \Tn with rounding
components, the last bit of the round-to-odd result in \Tlarge
captures the sticky bit. Similarly, the penultimate bit of the
round-to-odd result in \Tlarge captures the rounding bit.  In summary,
the round-to-odd result in \Tlarge maintains sufficient information
about the real value so that when the round-to-odd result is (double)
rounded to \Tsmall with any rounding mode, it produces the correctly
rounded result for \Tsmall.

\subsection{Polynomials for Correctly Rounded Results with the Round-to-Odd Mode in \Tlarge}
\input{alg.main.tex}
\input{alg.calcoddresult.tex}

Our strategy is to create a generic polynomial approximation that
produces correctly rounded results for \Tlarge using the round-to-odd
mode. Next, we describe our approach to generate such a polynomial
approximation. Algorithm~\ref{alg:main} provides a high-level sketch
of this process.  Given an elementary function $f(x)$ and a list of
inputs $X$ in the \Tn representation (\ie, $X \subseteq \Tn$), the
first step is to compute the correctly rounded result in
representation \Tlarge using the round-to-odd mode (\ie, $y_{\RNO}$
for each input $x \in X$). Figure~\ref{alg:calculateoddresult} shows
our algorithm to compute the round-to-odd result $y_{\RNO}$ for each
input using the real value from the oracle.

Subsequently, we compute the odd interval of each result $y_{\RNO}$
such that any real value in the odd interval rounds to
$y_{\RNO}$. Figure~\ref{alg:calculateoddresult} also provides our
algorithm to compute the odd interval. The odd intervals of some
inputs can be a singleton (\ie, only one value in the odd interval),
which we handle
separately~(Section~\ref{sec:approach:singleton}). Once we have a set
of non-singleton odd intervals for all inputs, we use our prior work
in the \rlibm project~\cite{lim:rlibm32:pldi:2021} to generate
piecewise polynomials.

At the end of this process, we will have two main components that
together can produce correctly rounded results for $f(x)$. First, our
approach produces a set $S$ that contains inputs whose odd interval is
a singleton.  For the resulting math libraries to be efficient, we
need a fast method to check these inputs and compute results for them
either using table lookups or using function-specific mathematical
properties~(Section~\ref{sec:approach:singleton}).  Second, our
approach produces piecewise polynomials that when used with output
compensation produces correct results for all inputs when rounded to
any \Tsmall with all standard rounding modes.

\textbf{Computing the round-to-odd result from a real value.} The
first step in our approach is to identify the correctly rounded result
$y_{\RNO}$ for input $x$.  Figure~\ref{alg:calculateoddresult}
provides the steps to compute the round-to-odd result in \Tlarge given
a real value of $f(x)$. We compute the real value $y = f(x)$ for each
input $x$ using an oracle (\eg, MPFR library). Then, we obtain the
rounding components ($s$, \Prec, \Rbit, \Sticky) as described in
Section~\ref{sec:background:rounding}. When the real value is exactly
representable (\Rbit =0 and \Sticky = 0) or when \Prec is odd, then
the round-to-odd result is \Prec. Otherwise, the round-to-odd result
in \Tlarge is the value succeeding \Prec in \Tlarge.

\textbf{Deducing the odd interval of an input.}  Once we determine the
correctly rounded result $y_{\RNO}$ of $f(x)$ in representation
\Tlarge using the round-to-odd mode, the next step is compute the
interval of values in representation $\mathbb{H}$, which is used for
polynomial evaluation and range reduction, such that producing any
value in the interval rounds to $y_{\RNO}$, which we call as the odd
interval. The function \texttt{CalcOddIntervals} in
Figure~\ref{alg:calculateoddresult} describes the steps to compute the
odd interval. If the correct rounded result $y_{\RNO}$ in \Tlarge is
even, then the odd interval is a singleton. In such cases, the only
value that rounds to $y_{\RNO}$ with the round-to-odd mode is
$y_{\RNO}$ itself.

Generating polynomial approximations with singletons is challenging
because they limit the amount of freedom available to the polynomial
generator. Hence, we identify such inputs and handle them
separately. If $y_{\RNO}$ is odd, then all values in $\mathbb{H}$ that
are strictly greater than the preceding value of $y_{\RNO}$ in \Tlarge
and strictly less than the succeeding value of $y_{\RNO}$ in \Tlarge
forms the odd interval. Any value in this odd interval rounds to
$y_{\RNO}$ in \Tlarge with the round-to-odd mode. We deduce the odd
interval for each input. In Figure~\ref{alg:calculateoddresult}, $L$
represents the set of non-singleton odd intervals for all inputs,
which is given to the polynomial generator.

\textbf{Piecewise polynomial generation using the odd intervals.} The
next step is to generate piecewise polynomials that produce a value in
the odd interval for all inputs. Each input and odd interval pair
(\ie, $(x, [l, h]) \in L$ specifies the constraints on the polynomial
approximation $A_{\mathbb{H}}(x)$ for each input $x$.  Using the
\rlibm methodology, we create an LP problem with these constraints to
deduce the coefficients of a polynomial with degree $d$. Similarly, we
generate piecewise polynomials and use counterexample guided
polynomial generation to facilitate the entire process. We also make
sure that the generated polynomial produces a value in the odd
interval after range reduction and output compensation.

\textbf{Implementation of the polynomial approximation for \Tsmall.}
At the end of polynomial generation, we will have a a set of inputs
whose odd interval is a singleton and a polynomial approximation of
$f(x)$ that produces the correct round-to-odd result in \Tlarge for
all inputs.  We implement the polynomial approximation as follows.
Given an input $x$, we first check whether the input $x$'s odd
interval is a singleton. If so, we either use the precomputed
round-to-odd result with table lookups or efficiently compute the
round-to-odd result using function-specific properties.  Otherwise, we
use perform range reduction and use Horner's method for polynomial
evaluation to compute the round-to-odd result in \Tlarge for the
input.  Finally, we round the round-to-odd result in \Tlarge to
\Tsmall using the user specified rounding mode to return the final
result. We guarantee that our implementation produces the correctly
rounded result of $f(x)$ for any representation \Tsmall with all the
standard rounding modes for all inputs $x$.

\subsection{Computing Round-to-Odd Results for Inputs with Singleton Odd Intervals}
\label{sec:approach:singleton}

One of the challenging issues for polynomial generation with odd
intervals is the presence of singletons, which happens when the
correctly rounded result with the round-to-odd mode is even (\ie, the
round-to-odd result in \Tlarge matches the real value).  We want to
identify such inputs efficiently.  As both \Tn and \Tlarge are finite
precision representations, all values in \Tn and \Tlarge are rational
values.  If the real value matches the round-to-odd result in \Tlarge
exactly, then it is a rational value. Hence, our task of identifying
inputs with singleton odd intervals corresponds to the problem of
identifying rational inputs that produce rational outputs for various
elementary functions, which is well
studied~\cite{niven:irrational:book:1956, Aigner:thebook:2009,
  baker:transcendental:1975, cohn:algebra:1974}.
We first use the mathematical properties of the elementary function
$f(x)$ to identify all rational inputs such that $f(x)$ is a rational
value.  Then, we check if these inputs $x_1$ and the corresponding
result $f(x_1)$ are exactly representable in \Tn and \Tlarge,
respectively. If so, such inputs are of interest. Then, we need to
develop a quick way to identify those inputs and compute the
round-to-odd results for them without using a multi-way branch.

We now describe the specific mathematical properties of elementary
functions that we use to identify inputs whose odd interval is a
singleton and the mechanism that we use to efficiently compute
round-to-odd results for them.

\textbf{Functions $e^{x}$ and $ln(x)$.} From the Lindemann-Weierstrass
theorem~\cite{baker:transcendental:1975}, if the input $x$ is a
non-zero rational value, then $e^{x}$ cannot be a rational
value. Hence, the only value that will have a singleton odd interval
with $e^x$ is $x=0$. Similarly, $ln(x)$ will produce a rational output
only when $x=1$, which also follows from the Lindemann-Weierstrass
theorem~\cite{baker:transcendental:1975}. We have a single branch to
check this input and return the pre-computed correctly rounded result.

\textbf{Functions $2^x$ and $10^x$.} The function $2^x$ can produce a
rational result only when $x$ is an integer and the value of $2^x$ is
less than the dynamic range of the \Tlarge representation. When \Tn is
a 32-bit float, \Tlarge can represent all values of $2^{x}$ for $x$
between $-151 \leq x \leq 127$. Thus, any integer input between $-151$
and $127$ (279 inputs in total) can produce a singleton odd
interval. Hence, our implementation checks whether $x$ is an integer
within a certain bound (\ie, $-151 \leq x \leq 127$) and directly
computes the result, $2^{x}$, using bit-wise operations.

Similar to $2^{x}$, $10^x$ produces a rational value when $x$ is a
positive integer. In contrast to $2^x$, $10^x$ grows much faster and
there are a few inputs for which $10^x$ is exactly representable in
\Tlarge.
For a 32-bit float (\Tn), there are only 12 inputs ranging from $0$ to
$11$ that are exactly representable in a 34-bit float (\Tlarge). We
use a precomputed table to store the correct results for these 12
inputs and use a switch statement for it.

\textbf{Functions $log_2(x)$ and $log_{10}(x)$.}  The $log_2(x)$
function produces a rational result when $x$ is a power of 2 (\ie, $x
= 2^k$ and $k$ is an integer). We use bitwise operations to check if
the input is a power of two. In contrast to $log_2(x)$, $log_{10}(x)$
produces a rational result when $x$ is a positive power of 10 (\ie, $x
= 10 ^k$ and $k$ is a positive integer). This difference between
$log_2(x)$ and $log_{10}(x)$ is due to the fact that \Tn cannot
exactly represent negative powers of 10.  When we are generating a
polynomial to approximate the round-to-odd result with a 34-bit float
(\ie, \Tlarge), there are 11 inputs that can produce singletons.  We
create table-lookups for them.

\textbf{The hyperbolic functions, $sinh(x)$ and $cosh(x)$.}  If the
input $x$ is a non-zero rational value, then $y = sinh(x)$ or
$y=cosh(x)$ cannot be a rational value using the Lindemann-Weierstrass
theorem~\cite{niven:irrational:book:1956}. Hence, the only input whose
odd interval is a singleton is $0$, for which we use a branch
condition.

\textbf{The $sinpi(x)$ function}. The function $sinpi(x)$ is equal to
$sin(\pi x)$. By Niven's theorem~\cite{niven:irrational:book:1956},
the only rational values of $x$ between $0 \leq x \leq \frac{1}{2}$
where $sinpi(x)$ is also a rational value are when $x = 0$,
$x=\frac{1}{6}$, and $x=\frac{1}{2}$. Among these three inputs,
$\frac{1}{6}$ is not exactly representable in \Tn. Given that $sinpi$
is a periodic function, there are only three cases of inputs in $x \in
\Tn$ where the result of $sinpi(x)$ is representable in \Tlarge when
we extend the domain of x to the set of all inputs:

\[
\small
sinpi(x) = 
\begin{cases}
0 & \text{if } x \text{ is an integer} \\
1 & \text{if } x \equiv \frac{1}{2} \mod 2.0 \\
-1 & \text{if } x \equiv \frac{3}{2} \mod 2.0
\end{cases}
\]

We need to implement the floating point modulo operation efficiently
using integer operations. Consider the case where \Tn is a 32-bit
float.  All inputs $x \in \Tn$ greater than or equal to $2^{23}$ are
integers where $sinpi(x)$ is always 0. Next, if $x < 2^{23}$, then we
need to identify whether $x$ is either an integer, a multiple of 0.5,
or a multiple of 1.5. To determine this condition, we compute $2x$
with a 32-bit float and then cast the result (\ie, $2x$) to a 32-bit
integer to obtain the value $t$. This operation of casting the value
$2x$ to an integer truncates the value of $2x$ to the integral part of
$2x$.  Now if we cast $t$ back to a 32-bit float value and the
resulting float value is exactly equal to $2x$, then $t$ is an
integer, which implies that $x$ is either an integer, or a multiple of
0.5, or a multiple of 1.5. Finally, we compute the result of
$sinpi(x)$ based on $t$ as shown below:

\[
\small
sinpi(x) = 
\begin{cases}
0 & \text{if } t \equiv 0 \mod 2 \\
1 & \text{if } t \equiv 1 \mod 4 \\
-1 & \text{if } t \equiv 3 \mod 4
\end{cases}
\]

\textbf{The $cospi(x)$ function.} Similarly, $cospi(x) = cos(\pi x)$ produces a
rational value representable in \Tlarge in the following cases.

\[
cospi(x) =
\begin{cases}
1 & \text{if } x \text{ is an even integer} \\
-1 & \text{if } x \text{ is an odd integer} \\
0 & \text{if } fraction(x) \equiv 0.5 
\end{cases}
\]

These checks can be performed efficiently using a similar strategy
illustrated for the $sinpi(x)$ function. In summary, handling
singleton odd intervals efficiently is important for performance when
we generate polynomials for correctly rounded results in \Tlarge with
the round-to-odd mode.

%% file: fig.rno_intuition.tex
\begin{figure}[!t]%
	\centering
	\includegraphics[width=\textwidth]{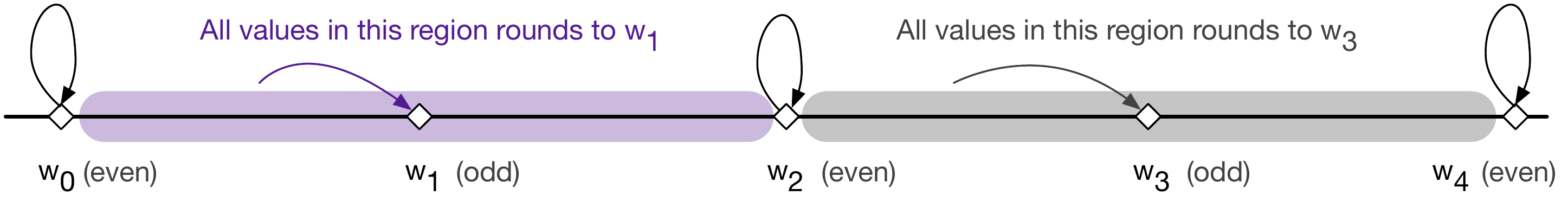}
	\caption{ The round-to-odd (\RNO) rounding mode. We show the
          rounding of \Real with the \RNO mode.  Here, $w_0$, $w_1$,
          $w_2$, $w_3$, and $w_4$ are values representable in
          representation $\mathbb{T}$. If \Real is exactly
          representable in $\mathbb{T}$, then \Real rounds to that
          value. Otherwise, \Real rounds to the nearest value in
          $\mathbb{T}$ that is odd.}
	\label{fig:rno_intuition}
\end{figure}

%% file: fig.rno_to_w1.tex
\begin{figure}[!t]%
	\centering
	\includegraphics[width=\textwidth]{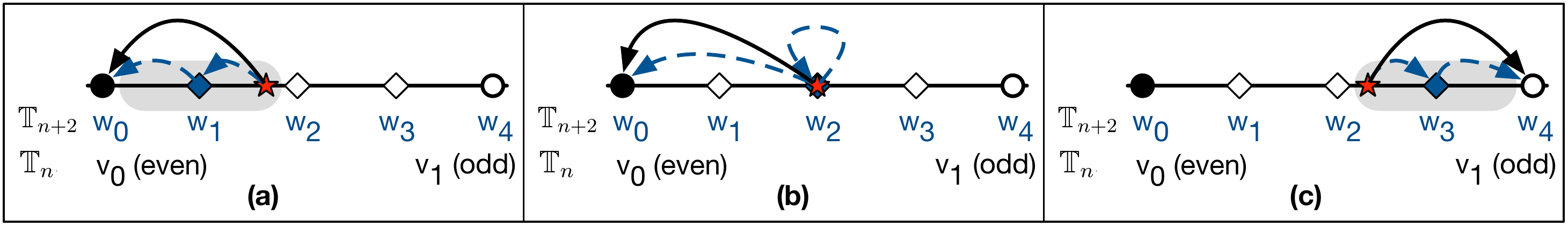}
	\caption{An example to show that the round-to-odd result in
          \Tlarge maintains sufficient information to produce
          correctly rounded results for \Tn when the round-to-odd
          result is double rounded to \Tn with the \RNE mode. Real
          value is represented with a red star. Here, $w_0$, $w_1$,
          $w_2$, $w_3$, and $w_4$ are representable values in
          \Tlarge. Values $v_0$ and $v_1$ are adjacent values
          representable in \Tn. As all values in \Tn are representable
          in \Tlarge, $w_0= v_0$ and $w_4 = v_1$. Solid arrow
          represents directly rounding the real value to \Tn. The
          dotted arrows represent the process of double rounding from
          the real value to the round-to-odd result in \Tlarge and
          subsequently to \Tn.  (a) When the real value is in the
          interval between $w_0$ and $w_2$. (b) When the real value is
          exactly equal to $w_2$, which is the midpoint of $v_0$ and
          $v_1$. (c) When the real value is between $w_2$ and $w_4$.}
	\label{fig:rno_to_w1}
\end{figure}

%% file: alg.main.tex
\begin{algorithm}[!t]
	\small
      \DontPrintSemicolon
      \SetKwFunction{FMain}{GenerateGenericPolynomial}
      \SetKwFunction{FCalculateL}{CalcOddIntervals}
      \SetKwFunction{FCalculateLp}{CalcReducIntervalsMulti}
      \SetKwFunction{FCalculateLambda}{CombineReducIntervals}
      \SetKwFunction{FGenPiecewisePoly}{GenPiecewisePoly}
      \SetKwFunction{FRLibmPolyGen}{RLibmPolyGen}
      \SetKwFunction{FCalculateOddResult}{CalcResultsInRO}
      \SetKwProg{Fn}{Function}{:}{}
      \Fn{\FMain{$f$, \Tlarge, $\mathbb{H}$, $X$, $d$, $RR_{\mathbb{H}}$, $OC_{\mathbb{H}}$}}{
        $O \leftarrow $ \FCalculateOddResult{$f$, \Tlarge, $X$}\;
      
        $(L, S) \leftarrow$ \FCalculateL{O, \Tlarge, $\mathbb{H}$}\;
        \lIf{$L = \emptyset$}{
          \Return{(false, $\emptyset$, DNE)}
        }
        $(status, P) \leftarrow $ \FRLibmPolyGen{$L$, $\mathbb{H}$, $d$, $RR_{\mathbb{H}}$, $OC_{\mathbb{H}}$}\;
        \Return{$(status, S, P)$}\;
      }
	\caption{\small A sketch of our approach to generate piecewise
          polynomials of degree $d$ for elementary function $f(x)$ in
          the representation \Tlarge using the round-to-odd mode.  The
          resulting polynomial when used with range reduction
          ($RR_{\mathbb{H}}$) and output compensation
          ($OC_{\mathbb{H}}$) produces correctly rounded results for
          all inputs $x \in X$ with all representations \Tsmall for
          all standard rounding modes. \texttt{CalcResultsInRO}
          computes the round-to-odd result using an oracle (see
          Figure~\ref{alg:calculateoddresult}). \texttt{CalcOddIntervals}
          computes the set of odd intervals~($L$) and set ($S$) of
          singleton odd intervals (see
          Figure~\ref{alg:calculateoddresult}). Once we have the odd
          intervals and singletons, we use \rlibm's polynomial
          generation procedure (\texttt{RLibmPolyGen}) to obtain the
          generic polynomial.}
	\label{alg:main}	
\end{algorithm}

%% file: alg.calcoddresult.tex
\begin{figure}
\small
\begin{subfigure}[t]{0.51\textwidth}
\begin{algorithm}[H]
      \DontPrintSemicolon
      \SetKwFunction{FMain}{GenerateGenericPolynomial}
      \SetKwFunction{FCalculateL}{CalcGenericIntervals}
      \SetKwFunction{FCalculateLp}{CalcReducIntervalsMulti}
      \SetKwFunction{FCalculateLambda}{CombineReducIntervals}
      \SetKwFunction{FGenPiecewisePoly}{GenPiecewisePoly}
      \SetKwFunction{FCalculateOddResult}{CalcResultsInRO}
      \SetKwFunction{FEncode}{EncodeBits}
      \SetKwFunction{FGetRComp}{RComp}
      \SetKwFunction{FIsOdd}{IsOdd}
      \SetKwFunction{FDecode}{DecodeBits}
      \SetKwFunction{FTrunc}{Trunc}
      \SetKwFunction{FConcat}{Concat}
      \SetKwFunction{FRedor}{ReduxOr}
	\SetKwFunction{FAdjHigher}{GetSuccVal}
      \SetKwProg{Fn}{Function}{:}{}
      \Fn{\FCalculateOddResult{$f$, \Tlarge, $X$}}{
      	$O \leftarrow \emptyset$\;
        \ForEach{$x \in X$} {
              $y$  = $f(x)$\;
		$(s, \Prec, rb, sticky) \leftarrow $ \FGetRComp{$y$, \Tlarge}\;
		\If{\FIsOdd{\Prec}$\lor (rb = 0 \land sticky = 0)$}{
			$y_{\RNO} \leftarrow s \times \Prec$
		}
		\Else{
			$\Succ \leftarrow$ \FAdjHigher{\Prec, \Tlarge}\;
			$y_{\RNO} \leftarrow s \times \Succ$\;
		}
		$O \leftarrow O \cup (x, y_{\RNO})$\;
        }
        \Return{$O$}
      }
\end{algorithm}
\end{subfigure}
\begin{subfigure}[t]{0.48\textwidth}
\begin{algorithm}[H]
      	\DontPrintSemicolon
	\SetKwFunction{FMain}{GenerateGenericPolynomial}
      	\SetKwFunction{FCalculateL}{CalcOddIntervals}
      	\SetKwFunction{FCalculateLp}{CalcReducIntervalsMulti}
      	\SetKwFunction{FCalculateLambda}{CombineReducIntervals}
      	\SetKwFunction{FGenPiecewisePoly}{GenPiecewisePoly}
      	\SetKwFunction{FCalculateOddResult}{CalcResultsInRO}
      	\SetKwFunction{FEncode}{EncodeBits}
	\SetKwFunction{FIsEven}{IsEven}
      	\SetKwFunction{FDecode}{DecodeBits}
      	\SetKwFunction{FTrunc}{Trunc}
      	\SetKwFunction{FConcat}{Concat}
      	\SetKwFunction{FRedor}{ReduxOr}
	\SetKwFunction{FAdjLower}{GetPrecVal}
	\SetKwFunction{FAdjHigher}{GetSuccVal}
      
      	\SetKwProg{Fn}{Function}{:}{}
      	\Fn{\FCalculateL{$O$, \Tlarge, $\mathbb{H}$}}{
		\ForEach{$(x, y_{\RNO}) \in O$} {
			$L \leftarrow \emptyset$\;
			$S \leftarrow \emptyset$\;
			\If{\FIsEven{$y_{\RNO}$}}{
				$S \leftarrow S \cup (x, y_{\RNO})$
			}
			\Else{
				$y^{-} \leftarrow$ \FAdjLower{$y_{\RNO}$, \Tlarge}\;
				$l \leftarrow$ \FAdjHigher{$y^{-}$, $\mathbb{H}$}\;
				$y^{+} \leftarrow$ \FAdjHigher{$y_{\RNO}$, \Tlarge}\;
				$h \leftarrow$ \FAdjLower{$y^{+}$, $\mathbb{H}$}\;
				$L \leftarrow L \cup (x, [l, h])$\;
			}
			\Return{$(L, S)$}
		}
      	}
\end{algorithm}
\end{subfigure}
\caption{\small \texttt{CalcResultsInRO} computes the correctly
  rounded result of $f(x)$ in \Tlarge using the round-to-odd rounding mode for
  each input $x \in X$. \texttt{CalcOddIntervals} computes the odd
  interval for each input $x$ based on the correctly rounded result
  $y_{\RNO}$ in \Tlarge.  The list $S$ is the set of inputs that have
  a singleton as the odd interval.  The list $L$ contains inputs and
  the corresponding odd intervals. \texttt{GetPrecVal}($a$,
  $\mathbb{T}$) returns the value preceding $a$ in the representation
  $\mathbb{T}$. \texttt{GetSuccVal}($a$, $\mathbb{T}$) returns the
  value succeeding $a$ in the representation $\mathbb{T}$.}
\label{alg:calculateoddresult}
\end{figure}

%% file: sec.new-proof.tex
\section{Proof that the Round-to-Odd Result with \Tlarge Produces Correct Results for \Tsmall}
\label{sec:rno:proof}

We provide a proof of Theorem~\ref{thm:main} in this section. We prove
that the round-to-odd result in \Tlarge produced by our polynomial
approximation when rounded to \Tsmall with any of the standard
rounding modes produces the correctly rounded result for \Tsmall.

\subsection{Unique Properties of the Round-to-Odd Result}

We prove the unique properties of the round-to-odd result, which we
subsequently use to prove Theorem~\ref{thm:main}. When we use \Real to
represent the real value, we refer to it in the extended infinite
precision representation.

\begin{lemma}{}
  \label{lemma:preserves_sign}
  The round-to-odd result \Vrno in \Tlarge preserves the sign of
  \Real.
\end{lemma}
\textbf{Proof.}
The value zero is representable in \Tlarge. The only value that rounds
to zero is zero itself. Hence, all positive real values will round to
a positive value in the round-to-odd mode. Similarly all negative real
values will round to a negative value in the round-to-odd mode.
$\qed$.

\begin{lemma}{}
\label{lemma:same_nbits_fp}
Let $\Vrno = \Round{\Tlarge}{\RNO}{\Real}$. The first $(n+1)$-bits of
\Vrno and \Real are identical.
\end{lemma}

\textbf{Proof.} The \Vrno result is created using the rounding
components (\svrno, \vmvrno, \rbvrno, \stickyvrno). The round-to-odd
mode preserves the sign of \Real in \Vrno. Without loss of generality,
we assume \Real is positive for the rest of the proof. Further,
\vmvrno is the truncated value of \Real (see
Section~\ref{sec:background:rounding}). Hence, all the $(n+2)$-bits of
\vmvrno and \Real are identical. After rounding with the round-to-odd
mode, \Vrno is either equal to \vmvrno or the succeeding value of
\vmvrno in \Tlarge.  We prove that the $(n+1)$-bits of \Vrno and \Real
are identical by looking at the possible values of \vmvrno and its
relation to \Vrno.

First case, when \vmvrno is odd. Then, \Vrno = \vmvrno. Hence, all the
$(n+2)$-bits of \Vrno and \Real are identical. Second case, when
\vmvrno is even. Hence, the last bit of \vmvrno is 0. Now, there are
two sub-cases.  (1) If \rbvrno $= 0$ and \stickyvrno $= 0$, then \Vrno
= \vmvrno. Hence, all the $(n+2)$-bits of \Vrno and \Real are
identical.  (2) If \rbvrno $\neq 0$ or \stickyvrno $\neq 0$, then
\Vrno is equal the succeeding value of \vmvrno. The only bit that
changes between \vmvrno and its succeeding value is the
$(n+2)^{th}$-bit. Hence, the first $(n+1)$-bits of \Vrno and \Real are
identical.  $\qed$.

\begin{lemma}{}
  \label{lemma:or_bits_fp}
  The ($n+2)^{th}$-bit of \Vrno is equal to the bitwise \texttt{OR} of
  all the bits of \Real starting from the $(n+2)^{th}$-bit.
\end{lemma}

\textbf{Proof.} We prove this lemma using a strategy similar to
Lemma~\ref{lemma:same_nbits_fp}. Intuitively, this lemma states that
the last bit of \Vrno is 0 if and only all bits starting from the
$(n+2)^{th}$-bit of \Real is 0.

As the round-to-odd mode preserves sign, we assume \Real is positive
for the rest of the proof without loss of generality. Let us say the
rounding components for \Vrno are (\svrno, \vmvrno, \rbvrno,
\stickyvrno). In the round-to-odd mode with \Tlarge, \Vrno will be
equal to either \vmvrno or a succeeding value of \vmvrno in \Tlarge.
Now, we look at the cases where \vmvrno is odd and even to complete
the proof.

In the first case, \vmvrno is odd. Then, \Vrno = \vmvrno. The
$(n+2)^{th}$-bit of \vmvrno is 1. As \vmvrno is a truncated value of
\Real, the $(n+2)^{th}$-bit of \Real is 1. Hence, the
bitwise-\texttt{OR} of all bits of \Real starting from the
$(n+2)^{th}$-bit is 1, which is equal to the $(n+2)^{th}$-bit of \Vrno
in \Tlarge.

In the second case, \vmvrno is even. There are two cases depending on
the values of \rbvrno and \stickyvrno. In the first sub-case, \rbvrno
$= 0$ and \stickyvrno = 0, then \Vrno = \vmvrno. The $(n+2)^{th}$-bit
of \vmvrno is 0. So is the $(n+2)^{th}$-bit of \Real. From the
definition of rounding components for \Tlarge, \rbvrno is the value of
the bit at position $(n+3)$ in \Real and \stickyvrno is the
bitwise-\texttt{OR} of bits starting from $(n+4)^{th}$-bit in
\Real. Hence, all the bits of \Real starting from the $(n+2)^{th}$-bit
are 0, which matches the $(n+2)^{th}$-bit of \Vrno.

The next sub-case is when \rbvrno $\neq 0$ or \stickyvrno $\neq 0$. In
this case, \Vrno is equal to the succeeding value of \vmvrno in
\Tlarge, which is odd. Hence, the $(n+2)^{th}$-bit of \Vrno is 1.
Both \stickyvrno and \rbvrno are not zeros, one of the bits starting
from $(n+2)^{th}$-bit in \Real is 1. Hence, the bitwise-\texttt{OR} of
all bits starting from the $(n+2)^{th}$-bit is 1, which matches the
$(n+2)^{th}$-bit of \Vrno.  $\qed$.

\begin{lemma}{}
\label{lemma:same_rounding_components}
Let ($s_1$, $\Prec_1$, $\Rbit_1$, $\Sticky_1$) and ($s_2$, $\Prec_2$,
$\Rbit_2$, $\Sticky_2$) be the rounding components for two real values
$v_1$ and $v_2$ in rounding them to a FP representation \Tn. If $s_1 =
s_2$, $\Prec_1 = \Prec_2$, $\Rbit_1 = \Rbit_2$, and $\Sticky_1 =
\Sticky_2$, then $\Round{\T}{rm}{v_1} = \Round{\T}{rm}{v_2}$ for any
rounding mode $rm$.
\end{lemma}

\textbf{Proof.} This lemma directly follows from the definition of rounding components
in Section~\ref{sec:background:rounding}. Intuitively, this lemma
states that identifying the correctly rounded result of the real value
in \Tn only depends on the rounding components and the rounding mode
$rm$. $\qed$.

\subsection{Proof that Double Rounding the Round-to-Odd Result Produces
  Correct Results for all \Tsmall}

We now sketch the proof of Theorem~\ref{thm:main}. We show that
rounding a real value \Real to the FP
representation $\Tlarge = \mathbb{F}_{n+2, |E|}$ using the
round-to-odd mode to produce \Vrno and then subsequently
rounding the result (\Vrno) to $\Tsmall$ using a rounding mode $rm$
produces the same value as rounding \Real directly to \Tsmall using
the same rounding mode $rm$, as long as $|E| + 1< k \leq n$. More
formally, we prove that

\[
\small
\Round{\Tsmall}{rm}{\Round{\Tlarge}{\RNO}{\Real}} = \Round{\Tsmall}{rm}{\Real}
\]

Our high-level strategy is to show that the rounding components for \Real
to \Tsmall and rounding \Vrno to \Tsmall are exactly the same. We
prove Theorem~\ref{thm:main} by proving the following theorem.

\input{fig.rno_to_fp.tex}

\begin{theorem}
\label{thm:rno_to_tp}
Given a real number \Real, representations \Tsmall and \Tlarge with
same number of exponent bits that satisfy the condition $|E| +1 < k
\leq n$, and a rounding mode $rm \in$ \{\RNE, \RNA, \RNZ, \RNP, \RNN
\}, then $\Round{\Tsmall}{rm}{\Real} =
\Round{\Tsmall}{rm}{\Round{\Tlarge}{\RNO}{\Real}}$.
\end{theorem}

Let us say $\Vrno = RN_{\Tlarge, \RNO}(\Real)$. Our goal is to prove
$RN_{\Tsmall, rm}(\Real) = RN_{\Tsmall, rm}(\Vrno)$. Using
Lemma~\ref{lemma:same_rounding_components}, if the rounding components
for rounding $\Real$ to $\Tsmall$ is the same as the rounding
components for rounding $\Vrno$ to $\Tsmall$, then we prove that
$\Round{\Tsmall}{rm}{\Real} = \Round{\Tsmall}{rm}{\Vrno}$ for all
rounding modes $rm$. Hence, our strategy is to show that the rounding
components for rounding $\Real$ and $\Vrno$ to $\Tsmall$ are
identical.  As the the round-to-odd mode preserves the sign, we will
consider \Real to be positive in the rest of the proof.

The representation of \Real in extended infinite precision
representation ($B_{\Real}$) is as follows:

\[
B_{\Real} = b_1 b_2 \dots b_{k-1} b_{k} b_{k+1} b_{k+2} \dots b_{n} b_{n+1} b_{n+2} b_{n+3} \dots
\]

Let us say ($s_1$, $\Prec_1$, $\Rbit_1$, $\Sticky_1$) are the rounding
components for rounding \Real to \Tsmall.  Then, $\Prec_1$ is the
truncated value in \Tsmall. While rounding to \Tsmall, the rounding
bit, $\Rbit_1$, is the $(k+1)^{th}$-bit of \Real. The sticky bit,
$\Sticky_1$ is the bitwise-\texttt{OR} of all bits starting from the
$(k+2)^{th}$-bit of \Real.

\[
B_{\Prec_1} = b_1 b_2 b_3 \dots b_{k-1} b_k, \quad\quad \Rbit_1 = b_{k+1}, \quad\quad \Sticky_1 = b_{k+2} \:|\: b_{k+3} \:|\: \dots \nonumber
\]

Figure~\ref{fig:rno_to_fp} pictorially shows the rounding
components $\Prec_1$, $\Rbit_1$, and $\Sticky_1$ while rounding \Real to \Tsmall.

Similarly, we next identify the rounding components ($s_2$, $\Prec_2$,
$\Rbit_2$, $\Sticky_2$) for rounding \Vrno to \Tsmall. Note that \Vrno
is a result in \Tlarge. From Lemma~\ref{lemma:same_nbits_fp} and
Lemma~\ref{lemma:or_bits_fp}, the bit-string of \Vrno is:

\[
\small
B_{\Vrno} = b_1 b_2 b_3 \dots b_{k-1} b_{k} \dots b_n b_{n+1} t, \quad \quad  t = b_{n+2} \mid b_{n+3} \mid b_{n+4} \mid \dots \nonumber
\]

Since $k \leq n$, there are at least one bit (\ie, $b_{n+1}$) between
$b_k$ and $t$, where $t$ is the $(n+2)^{th}$-bit in \Vrno.
The rounding components when we round \Vrno to \Tsmall are:

\[
\small
B_{\Prec_2} = b_1 b_2 b_3 \dots b_{k-1} b_k, \quad\quad \Rbit_2 = b_{k+1} \quad\quad \Sticky_2 = b_{k+2} \mid b_{k+3} \mid \dots \mid b_{n+1} \mid t
\]

Figure~\ref{fig:rno_to_fp} shows these components while rounding \Vrno
to \Tsmall.

Now, we compare the rounding components when we directly round \Real
to \Tsmall with the rounding components when we round \Vrno to
\Tsmall.  The sign information ($s_1$ and $s_2$) is identical because
the round-to-odd mode preserves the sign of \Real. The truncated values,
$\Prec_1$ and $\Prec_2$, are equal because their bit-strings are
identical. The rounding bit, $\Rbit_1$ and $\Rbit_2$, is identical and
is equal to $b_{k+1}$. Let us look at the sticky bits, $\Sticky_1$ and
$\Sticky_2$:

\begin{align}
\small
\Sticky_2 &=  b_{k+2} \mid b_{k+3} \mid \dots \mid b_{n+1} \mid t = b_{k+2} \mid b_{k+3} \mid \dots \mid b_{n+1} \mid b_{n+2} \mid b_{n+3} \mid b_{n+4} \mid \dots \nonumber  \\ 
&= \Sticky_1 \nonumber \\
\end{align}

Hence, all the rounding components for rounding \Real to \Tsmall
directly and rounding \Vrno to \Tsmall are identical. Hence,
$\Round{\Tsmall}{rm}{\Real} = \Round{\Tsmall}{rm}{\Vrno}$ from
Lemma~\ref{lemma:same_rounding_components}.  $\qed$

Theorem~\ref{thm:main} directly follows from
Theorem~\ref{thm:rno_to_tp}. Theorem~\ref{thm:rno_to_tp} states that
\Vrno, which is produced by rounding a real value \Real to \Tlarge
using the round-to-odd mode, rounds to the same value as if \Real is
directly rounded to \Tsmall using the same rounding mode $rm$. If we
substitute \Real with the exact result of the elementary function
$f(x)$ for a given input $x \in \Tn$, then

\[
\small
\Round{\Tsmall}{rm}{\Round{\Tlarge}{\RNO}{f(x)}} = \Round{\Tsmall}{rm}{f(x)}
\]

Further, by definition, all values in the odd interval of \Vrno in
\Tlarge round to \Vrno. Hence, any value in the odd interval rounds to
the correctly rounded result for representations \Tsmall using any
rounding mode $rm \in$ \{\RNE, \RNA, \RNZ, \RNP, \RNN \}.

%% file: fig.rno_to_fp.tex
\begin{figure}[!t]%
	\centering
        \includegraphics[width=0.90\textwidth]{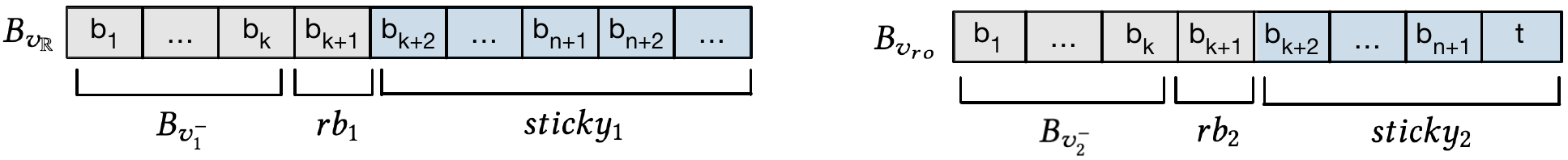}
	\caption{Rounding components while rounding \Real
          and \Vrno to \Tsmall. We show the bit-string of \Real in
          extended infinite precision representation. Note \Vrno is a
          value in \Tlarge.}
	\label{fig:rno_to_fp}
\end{figure}

%% file: sec.eval.tex
\section{Experimental Evaluation}
\label{sec:eval}
We describe our prototype, experimental methodology, and the results
of our experiments to check the correctness and performance of the
generated polynomial approximations.

\textbf{Prototype.} The prototype, \tool, is open-source and publicly
available~\cite{rlibm-all}. \tool is a generator and a collection of
correctly rounded implementations of polynomial approximation for
multiple representations and rounding modes.
\tool contains ten functions that produce the correctly rounded result
of $f(x)$ for the 34-bit FP representation (\ie, \Tlarge) with 8 bits
of exponent (FP34) with the round-to-odd mode. As FP34 is not
supported in hardware, \ourlibm maintains the FP34 result in double
precision. \tool's functions produce the correct result for all
$n$-bit FP representations with 8-bits for the exponent with all the
rounding modes in the IEEE standard where $9 < n \leq 32$. This
includes 32-bit float, bfloat16, and tensorfloat32. \tool uses the
MPFR library~\cite{Fousse:toms:2007:mpfr} with up to 1000 precision
bits to compute the oracle value of $f(x)$. \tool uses
SoPlex~\cite{Gleixner:soplex:issac:2012}, an exact rational LP solver,
to generate the coefficients of the polynomials with a time limit of
five minutes. We limit the size of the LP formulation to contain up to
fifty thousand reduced input and interval constraints. We use the
range reduction and output compensation functions from the \rlibm
prototype. \tool performs range reduction, polynomial evaluation, and
output compensation using the double precision. The polynomial
evaluation uses the Horner's
method~\cite{borwein:polynomials:book:1995} for efficiency.

\textbf{Experimental methodology and setup. }
We compare \tool's functions with Intel's libm, glibc's libm,
CR-LIBM~\cite{Daramy:crlibm:doc}, and \Rlibmtt.  Among these, CR-LIBM
provides four implementations for each elementary function that
produces the correctly rounded results in double precision with the
\RNE, \RNZ, \RNP, and \RNN mode, respectively. CR-LIBM does not
provide implementations for the \RNA mode. \Rlibmtt provides correctly
rounded functions for a 32-bit float with the \RNE mode.
To produce the result in a target representation \T that is not
natively supported by these libraries, we first convert the input in
\T to the representation supported by the library, use the elementary
function, and round the result back to \T.
We perform our experiments on a 2.10GHz Intel Xeon Gold 6230R machine
with 192GB of RAM running Ubuntu 18.04. We disabled Intel turbo boost
and hyper-threading to minimize noise. All our libraries are compiled
with \texttt{O3} optimizations. We use Intel's libm from the oneAPI
Toolkit and glibc's libm from glibc-2.33.  The test harness for
comparing glibc's libm, CR-LIBM, and \Rlibmtt is built using the
\texttt{gcc-10} compiler with \texttt{-O3 -static -frounding-math
  -fsignaling-nans} flags. Because Intel's libm is only supported in
the Intel's compiler, we built a test harness that compares Intel's
libm against \tool using the \texttt{icc} compiler with \texttt{-O3
  -static -no-ftz -fp-model strict} flags to obtain as many accurate
results as possible.
To compare performance, we measure the number of cycles taken to
compute the result for each input using \texttt{rdtscp}. We then
measured the total time taken to compute the elementary function as
the sum of the time taken by all inputs.

\input{tbl.eval.ofa.statistics.tex}

\subsection{Polynomial Generation with \tool}
Table~\ref{tbl:eval:ofa:statistics} provides details on the properties
of the polynomials generated by \tool. Our attempt was to generate
piecewise polynomials with degree less than or equal to 8. We also
restricted the number of sub-domains for the piecewise polynomials to
$2^{15}$. The output compensation function for $sinh(x)$, $cosh(x)$,
$sinpi(x)$, and cospi$(x)$ uses two elementary functions and we
generate two piecewise polynomials for each function. The $e^x$,
$2^x$, and $10^{x}$ functions have both negative and positive reduced
inputs. Hence, we create two piecewise polynomials: one for the
negative reduced inputs and the other for the positive reduced inputs.
As \ourlibm generates piecewise polynomials for FP34 with the
round-to-odd mode, the number of sub-domains used in the resulting
piecewise polynomials are bigger than \Rlibmtt. However, the degrees
of the polynomial for each sub-domain was similar to \Rlibmtt. The
amount of time taken to generate the piecewise polynomials ranged from
approximately 2 hours to 9 hours. About 79\% of the total time on
average is spent in computing the oracle result using the MPFR
library.  In contrast, computing the intervals and generating the
polynomials using the LP solver takes 15\% and 5\% of the total time
on average, respectively.

\subsection{Does \tool Produce Correct Results?}
We experimentally show that \tool produces correctly rounded results
for all rounding modes for multiple representations. We built a
harness that checks if \tool's functions produce correctly rounded
results for all inputs with 161 different FP representations where the
number of exponent bits ranged from $2$ to $8$ and the number of
mantissa bits ranged from $1$ to $23$ bits (\ie, $23 * 7 = 161$) for
all five rounding modes. \tool produces the correct results for all
these representations with all standard rounding modes with all
inputs.
\emph{\tool is the first efficient library that provides correctly
rounded results for all rounding modes for a 32-bit float.}

\input{tbl.eval.ofa.floatacc.tex}

\textbf{Correct results with all rounding modes for a 32-bit float.}
Table~\ref{tbl:eval:ofa:floatacc} reports the results of our
experiments to check whether existing libraries produce correct
results for a 32-bit float type. While \tool produces the correct
float results with all five rounding modes, mainstream libraries
(glibc and Intel's libm) do not produce correctly rounded results for
all rounding modes for all inputs for many of the elementary
functions. When CR-LIBM's \RNE implementations, which produces
correctly rounded double results with the \RNE mode for all inputs, is
used to produce the \RNE results for a 32-bit float, it does not
produce correctly rounded results for all inputs with several
functions due to double rounding. CR-LIBM's \RNN, \RNP, and \RNZ
implementations produce correctly rounded float results for the \RNN,
\RNP, and \RNZ mode respectively. Double rounding with these three
rounding modes using CR-LIBM's implementations do not generate wrong
results.

\Rlibmtt produces correctly rounded results for all inputs for the
\RNE mode with a 32-bit float. However, it does not produce correct
results for other rounding modes. In contrast, \ourlibm's produces a
single polynomial approximation for an elementary function that
produces correct results for all inputs and for all rounding modes.

\input{tbl.eval.ofa.tf32acc.tex}

\textbf{Correct results with all rounding modes for tensorfloat32.}
Tensorfloat32 is a new 19-bit representation with the same number of
exponents bits as a 32-bit float.  Tensorfloat32 has the same number
of exponent bits as \ourlibm's FP34.  Hence, \tool produces correctly
rounded results for all inputs and for all rounding modes with
tensorfloat32.  Table~\ref{tbl:eval:ofa:tf32acc} shows that glibc's
libm, Intel's libm, and RLibm32 do not produce correct results for all
rounding modes for all ten elementary functions. Although glibc's and
Intel's libm are designed to produce double results, which has
significantly higher precision than tensorfloat32, the double rounding
error still results in incorrect tensorfloat32 results for \RNN, \RNP,
and \RNZ rounding modes especially with extremal values. In contrast,
CR-LIBM's implementation designed for each rounding mode $rm$ produces
correctly rounded results for all inputs with the same rounding mode
$rm$. However, using CR-LIBM's implementation for a specific rounding
mode to produce the results for another rounding mode results in wrong
results. \tool is the first collection of elementary functions for
tensorfloat32 that produces correct results for all inputs and for all
rounding modes with a single polynomial approximation.

\input{tbl.eval.ofa.bf16acc.tex}

\textbf{Correct results with all rounding modes for bfloat16.} As
bfloat16 also has the same number of exponent bits as \ourlibm's FP34,
\tool produces correctly rounded results for all inputs and for all
rounding modes with it. Similar to tensorfloat32, mainstream libraries
do not produce correctly rounded results for all inputs and for all
rounding modes with bfloat16 due to double rounding issues.  In
contrast, \tool produces correct results for all bfloat16 inputs and
for all rounding modes with a single polynomial approximation.

\subsection{Performance Evaluation of \tool's Functions}
\input{fig.eval.ofa.GrlibmFloatSpeedup.tex}

Figure~\ref{fig:eval:ofa:GrlibmFloatSpeedup} reports the speedup of
\tool's functions over various mainstream libraries (glibc's libm and
Intel's libm) and correctly rounded libraries (CR-LIBM and \Rlibmtt).
Figure~\ref{fig:eval:ofa:GrlibmFloatSpeedup}(a) presents the speedup
of \tool's FP functions over glibc's float functions (left bar in each
cluster) and double functions (right bar in each cluster). On average,
\tool's FP functions are $1.05\times$ and $1.1\times$ faster than
glibc's float and double functions, respectively.
Figure~\ref{fig:eval:ofa:GrlibmFloatSpeedup}(b) presents the speedup
of \tool's FP functions over Intel's float functions (left bar in each
cluster) and double functions (right bar in each cluster). On average,
\tool has $1.34\times$ and $1.46\times$ speedup over Intel's float and
double functions,
respectively. Figure~\ref{fig:eval:ofa:GrlibmFloatSpeedup}(c) presents
the speedup of \tool's FP functions over CR-LIBM functions. On
average, \tool has $1.86\times$ speedup over CR-LIBM functions.  In
contrast to \tool, glibc's libm, Intel's libm, and CR-LIBM do not
produce correct results for all inputs when used for a 32-bit float
type.

Figure~\ref{fig:eval:ofa:GrlibmFloatSpeedup}(d) presents the speedup
of \tool's FP functions over \Rlibmtt's functions in producing 32-bit
float values rounded with the \RNE rounding mode. On average, \tool is
almost as fast as \Rlibmtt (\ie, 2\% slower than \Rlibmtt). \tool
creates polynomial approximations for a 34-bit FP representation,
which is the main reason for this small performance slowdown.
Notably, \tool experiences roughly a 12\% slowdown with $sinh(x)$ when
compared to \Rlibmtt.  When an input $x$ is near $0$, $sinh(x)$
exhibits a linear behavior and $sinh(x) \approx x$ produces correctly
rounded float values with the \RNE rounding mode. \Rlibmtt uses this
property by simply returning $x$ for inputs near $0$. In contrast,
\tool performs significantly more computation to ensure that it
produces a value within the odd interval.
Unlike \Rlibmtt that produces correct results for a single
representation with the \RNE mode, \tool produces correct results for
all inputs for multiple representations and all the standard rounding
modes.

%% file: tbl.eval.ofa.statistics.tex
\begin{table}[!t]
  \footnotesize
  \caption{Details about the generated polynomials. For each function,
    we show the time taken to generate the polynomial in minutes, the
    size of the piecewise polynomial, the maximum degree, the number
    of terms, and whether the generated polynomial produces correct
    results in FP34 using the round-to-odd mode for all inputs.}
  \begin{tabular}[t]{| c | c | c | c | c | c |}
    \hline
	$f(x)$
      & \begin{tabular}{@{}c@{}c@{c}}Gen. \\ Time \\ (Min.)\end{tabular}
      & \begin{tabular}{@{}c@{}}\# of Poly- \\ nomials \end{tabular}
      & \begin{tabular}{@{}c@{}}Deg-\\ree\end{tabular} 
      & \begin{tabular}{@{}c@{}}\# of \\ Terms\end{tabular}
      & \begin{tabular}{@{}c@{}}FP34 \\ \RNO \end{tabular}\\
      \hline
      $\mathbf{ln(x)}$  	& $325$ 		& $2^{10}$ & 3 & 3 & \cmark	\\ \hline
      $\mathbf{log_2(x)}$  	& $420$ 	& $2^{8}$ & 3 & 3 & \cmark	\\ \hline
      $\mathbf{log_{10}(x)}$  	& $546$ 	& $2^{8}$ & 3 & 3 & \cmark	\\ \hline
      $\mathbf{e^x}$  	& $241$ 			& 
      	\begin{tabular}{@{}c@{}}$2^{7}$ \\ $2^{7}$\end{tabular} & 
	\begin{tabular}{@{}c@{}} 4 \\ 4\end{tabular} & 
	\begin{tabular}{@{}c@{}} 5 \\ 5\end{tabular} & \cmark\\ \hline
      $\mathbf{2^x}$  	& $151$ 		& 
      	\begin{tabular}{@{}c@{}}$2^{7}$ \\ $2^{7}$\end{tabular}  & 
	\begin{tabular}{@{}c@{}} 3 \\ 3\end{tabular} & 
	\begin{tabular}{@{}c@{}} 4 \\ 4\end{tabular} & \cmark\\ \hline
   \end{tabular}
    \begin{tabular}[t]{| c | c | c | c | c | c |}
      \hline
	$f(x)$
      & \begin{tabular}{@{}c@{}c@{c}}Gen. \\ Time \\ (Min.)\end{tabular}
      & \begin{tabular}{@{}c@{}}\# of Poly- \\ nomials \end{tabular}
      & \begin{tabular}{@{}c@{}}Deg-\\ree\end{tabular} 
      & \begin{tabular}{@{}c@{}}\# of \\ Terms\end{tabular}
      & \begin{tabular}{@{}c@{}}FP34 \\ \RNO \end{tabular}\\
      \hline
      $\mathbf{10^x}$ 	& $402$ 		& 
      	\begin{tabular}{@{}c@{}}$2^{8}$ \\ $2^{8}$\end{tabular}  & 
	\begin{tabular}{@{}c@{}} 3 \\ 3\end{tabular} & 
	\begin{tabular}{@{}c@{}} 4 \\ 4\end{tabular} & \cmark\\ \hline
      $\mathbf{sinh(x)}$  	& $143$ 		 & $2^6$ & 5 & 3  & \cmark	\\  
      $\mathbf{cosh(x)}$  	& $135$ 		 & $2^5$ & 4 & 3 & \cmark	\\  \hline
      $\mathbf{sinpi(x)}$  	& $308$ 		 & $2^2$ & 5 & 3 & \cmark	\\ 
      $\mathbf{cospi(x)}$  	& $316$ 		 & $2^2$ & 4 & 3 & \cmark	\\ \hline
   \end{tabular}
\label{tbl:eval:ofa:statistics}
\end{table}

%% file: tbl.eval.ofa.floatacc.tex
\begin{table*}
	\footnotesize
	\caption{Ability to generate correct results for a 32-bit
          float for all inputs with each of the five standard rounding
          modes with \tool, glibc's double libm, Intel's double libm,
          CR-LIBM, and \Rlibmtt. CR-LIBM provides separate correctly
          rounded functions for each rounding mode except the $\RNA$
          mode, which we use to check its correctness.  \cmark
          indicates that the library produces the correct result using
          a given rounding mode for all inputs. Otherwise, we use
          \xmark.}
  
\begin{tabular}{| P{3.5em} | P{1.3em} | P{1.3em} | P{1.3em} | P{1.3em} | P{1.3em} |}
	\hline
	& \multicolumn{5}{| c |}{\textbf{Using \tool}} \\
	\hline
	\hline
	$f(x)$ & \RNE & \RNN & \RNP & \RNZ & \RNA\\
	\hline
	$\mathbf{ln(x)}$ & \cmark & \cmark & \cmark & \cmark & \cmark \\ \hline
	$\mathbf{log_2(x)}$ & \cmark & \cmark & \cmark & \cmark & \cmark  \\ \hline
	$\mathbf{log_{10}(x)}$ & \cmark & \cmark & \cmark & \cmark & \cmark  \\ \hline
	$\mathbf{e^{x}}$ & \cmark & \cmark & \cmark & \cmark & \cmark \\ \hline
	$\mathbf{2^{x}}$ & \cmark & \cmark & \cmark & \cmark & \cmark  \\ \hline
	$\mathbf{10^{x}}$ & \cmark & \cmark & \cmark& \cmark & \cmark  \\ \hline
	$\mathbf{sinh(x)}$ & \cmark & \cmark & \cmark& \cmark & \cmark \\  \hline
	$\mathbf{cosh(x)}$ & \cmark & \cmark & \cmark & \cmark & \cmark  \\  \hline
	$\mathbf{sinpi(x)}$ & \cmark & \cmark& \cmark & \cmark & \cmark  \\ \hline
	$\mathbf{cospi(x)}$ & \cmark & \cmark & \cmark & \cmark & \cmark  \\ \hline
\end{tabular}
\begin{tabular}{| P{1.3em} | P{1.3em} | P{1.3em} | P{1.3em} | P{1.3em} |}
	\hline
	\multicolumn{5}{| c |}{Using glibc double libm} \\
	\hline
	\hline
	\RNE & \RNN & \RNP & \RNZ & \RNA\\
	\hline
	\xmark & \xmark & \xmark & \xmark & \xmark \\ \hline
	\cmark & \cmark & \cmark & \cmark & \cmark \\ \hline
	\xmark & \xmark & \xmark & \xmark & \xmark \\ \hline
	\cmark & \xmark & \xmark & \xmark & \cmark \\ \hline
	\xmark & \xmark & \xmark & \xmark & \xmark \\ \hline
	\cmark & \xmark & \xmark & \xmark & \xmark \\ \hline
	\xmark & \xmark & \xmark & \xmark & \xmark \\ \hline
	\cmark & \xmark & \xmark & \xmark & \cmark \\ \hline
	\NA & \NA& \NA & \NA & \NA  \\ \hline
	\NA & \NA & \NA & \NA & \NA  \\ \hline
\end{tabular}
\begin{tabular}{| P{1.3em} | P{1.3em} | P{1.3em} | P{1.3em} | P{1.3em} |}
	\hline
	\multicolumn{5}{| c |}{Using Intel double libm} \\
	\hline
	\hline
	\RNE & \RNN & \RNP & \RNZ & \RNA\\
	\hline
	\xmark & \xmark & \xmark & \xmark & \xmark \\ \hline
	\cmark & \cmark & \cmark & \cmark & \cmark \\ \hline
	\xmark & \xmark & \xmark & \xmark & \xmark \\ \hline
	\cmark & \xmark & \xmark & \xmark & \cmark \\ \hline
	\xmark & \xmark & \xmark & \xmark & \xmark \\ \hline
	\cmark & \xmark & \xmark & \xmark & \cmark \\ \hline
	\xmark & \xmark & \xmark & \xmark & \xmark \\ \hline
	\cmark & \xmark & \xmark & \xmark & \cmark \\ \hline
	\cmark & \cmark & \cmark & \cmark & \xmark \\ \hline
	\cmark & \xmark & \cmark & \xmark & \xmark \\ \hline
\end{tabular}

\begin{tabular}{| P{3.5em} | P{1.3em} | P{1.3em} | P{1.3em} | P{1.3em} | P{1.3em} |}
	\hline
	& \multicolumn{5}{| c |}{Using CRLIBM} \\
	\hline
	\hline
	$f(x)$ & \RNE & \RNN & \RNP & \RNZ & \RNA\\
	\hline
	$\mathbf{ln(x)}$ & \xmark & \cmark & \cmark & \cmark & \NA \\ \hline
	$\mathbf{log_2(x)}$ & \cmark & \cmark & \cmark & \cmark & \NA \\ \hline
	$\mathbf{log_{10}(x)}$ & \xmark & \cmark & \cmark & \cmark & \NA \\ \hline
	$\mathbf{e^{x}}$ & \cmark & \cmark & \cmark & \cmark & \NA \\ \hline
	$\mathbf{2^{x}}$ & \NA & \NA & \NA & \NA & \NA \\ \hline
	$\mathbf{10^{x}}$ & \NA & \NA & \NA & \NA & \NA \\ \hline
	$\mathbf{sinh(x)}$ & \xmark & \cmark & \cmark & \cmark & \NA \\ \hline
	$\mathbf{cosh(x)}$ & \cmark & \cmark & \cmark & \cmark & \NA \\ \hline
	$\mathbf{sinpi(x)}$ & \cmark & \cmark & \cmark & \cmark & \NA \\ \hline
	$\mathbf{cospi(x)}$ & \cmark & \cmark & \cmark & \cmark & \NA \\ \hline
\end{tabular}
\begin{tabular}{| P{1.3em} | P{1.3em} | P{1.3em} | P{1.3em} | P{1.3em} |}
	\hline
	\multicolumn{5}{| c |}{Using \Rlibmtt} \\
	\hline
	\hline
	\RNE & \RNN & \RNP & \RNZ & \RNA\\
	\hline
	\cmark & \xmark & \xmark & \xmark & \cmark \\ \hline
	\cmark & \xmark & \xmark & \xmark & \cmark \\ \hline
	\cmark & \xmark & \xmark & \xmark & \cmark \\ \hline
	\cmark & \xmark & \xmark & \xmark & \cmark \\ \hline
	\cmark & \xmark & \xmark & \xmark & \xmark \\ \hline
	\cmark & \xmark & \xmark & \xmark & \cmark \\ \hline
	\cmark & \xmark & \xmark & \xmark & \cmark \\ \hline
	\cmark & \xmark & \xmark & \xmark & \cmark \\ \hline
	\cmark & \xmark & \xmark & \xmark & \cmark \\ \hline
	\cmark & \xmark & \xmark & \xmark & \cmark \\ \hline
\end{tabular}

\label{tbl:eval:ofa:floatacc}
\end{table*}

%% file: tbl.eval.ofa.tf32acc.tex
\begin{table*}
	\footnotesize
	\caption{Ability to generate correct results with
          tensorfloat32 for all inputs with various rounding
          modes. \cmark indicates that the library produces the
          correct tensorfloat32 result using a given rounding mode for
          all inputs. Otherwise, we use \xmark.}
  
\begin{tabular}{| P{3.5em} | P{1.3em} | P{1.3em} | P{1.3em} | P{1.3em} | P{1.3em} |}
	\hline
	& \multicolumn{5}{| c |}{\textbf{Using \tool}} \\
	\hline
	\hline
	$f(x)$ & \RNE & \RNN & \RNP & \RNZ & \RNA\\
	\hline
	$\mathbf{ln(x)}$ & \cmark & \cmark & \cmark & \cmark & \cmark \\ \hline
	$\mathbf{log_2(x)}$ & \cmark & \cmark & \cmark & \cmark & \cmark  \\ \hline
	$\mathbf{log_{10}(x)}$ & \cmark & \cmark & \cmark & \cmark & \cmark  \\ \hline
	$\mathbf{e^{x}}$ & \cmark & \cmark & \cmark & \cmark & \cmark \\ \hline
	$\mathbf{2^{x}}$ & \cmark & \cmark & \cmark & \cmark & \cmark  \\ \hline
	$\mathbf{10^{x}}$ & \cmark & \cmark & \cmark& \cmark & \cmark  \\ \hline
	$\mathbf{sinh(x)}$ & \cmark & \cmark & \cmark& \cmark & \cmark \\  \hline
	$\mathbf{cosh(x)}$ & \cmark & \cmark & \cmark & \cmark & \cmark  \\  \hline
	$\mathbf{sinpi(x)}$ & \cmark & \cmark& \cmark & \cmark & \cmark  \\ \hline
	$\mathbf{cospi(x)}$ & \cmark & \cmark & \cmark & \cmark & \cmark  \\ \hline
\end{tabular}
\begin{tabular}{| P{1.3em} | P{1.3em} | P{1.3em} | P{1.3em} | P{1.3em} |}
	\hline
	\multicolumn{5}{| c |}{Using glibc double libm} \\
	\hline
	\hline
	\RNE & \RNN & \RNP & \RNZ & \RNA\\
	\hline
	\cmark & \cmark & \cmark & \cmark & \cmark \\ \hline
	\cmark & \cmark & \cmark & \cmark & \cmark  \\ \hline
	\cmark & \cmark & \cmark & \cmark & \cmark  \\ \hline
	\cmark & \xmark & \xmark & \xmark & \cmark \\ \hline
	\cmark & \xmark & \xmark& \xmark & \cmark  \\ \hline
	\cmark & \xmark & \xmark & \xmark & \cmark  \\ \hline
	\cmark & \xmark & \xmark& \xmark & \cmark \\  \hline
	\cmark & \xmark & \xmark & \xmark & \cmark  \\  \hline
	\NA & \NA& \NA & \NA & \NA  \\ \hline
	\NA & \NA & \NA & \NA & \NA  \\ \hline
\end{tabular}
\begin{tabular}{| P{1.3em} | P{1.3em} | P{1.3em} | P{1.3em} | P{1.3em} |}
	\hline
	\multicolumn{5}{| c |}{Using Intel double libm} \\
	\hline
	\hline
	\RNE & \RNN & \RNP & \RNZ & \RNA\\
	\hline
	\cmark & \cmark & \cmark & \cmark & \cmark \\ \hline
	\cmark & \cmark & \cmark & \cmark & \cmark  \\ \hline
	\cmark & \cmark & \cmark & \cmark & \cmark  \\ \hline
	\cmark & \xmark & \xmark & \xmark & \cmark \\ \hline
	\cmark & \xmark & \xmark & \xmark & \cmark  \\ \hline
	\cmark & \xmark & \xmark& \xmark & \cmark  \\ \hline
	\cmark & \xmark & \xmark& \xmark & \cmark \\  \hline
	\cmark & \xmark & \xmark & \xmark & \cmark  \\  \hline
	\cmark & \cmark& \cmark & \cmark & \cmark  \\ \hline
	\cmark & \xmark & \cmark & \xmark & \cmark  \\ \hline
\end{tabular}

\begin{tabular}{| P{3.5em} | P{1.3em} | P{1.3em} | P{1.3em} | P{1.3em} | P{1.3em} |}
	\hline
	& \multicolumn{5}{| c |}{Using CRLIBM} \\
	\hline
	\hline
	$f(x)$ & \RNE & \RNN & \RNP & \RNZ & \RNA\\
	\hline
	$\mathbf{ln(x)}$ & \cmark & \cmark & \cmark & \cmark & \NA \\ \hline
	$\mathbf{log_2(x)}$ & \cmark & \cmark & \cmark & \cmark & \NA  \\ \hline
	$\mathbf{log_{10}(x)}$ &  \cmark & \cmark & \cmark & \cmark & \NA  \\ \hline
	$\mathbf{e^{x}}$ & \cmark & \cmark & \cmark & \cmark & \NA \\ \hline
	$\mathbf{2^{x}}$ & \NA & \NA & \NA & \NA & \NA  \\ \hline
	$\mathbf{10^{x}}$ & \NA & \NA & \NA& \NA & \NA  \\ \hline
	$\mathbf{sinh(x)}$ & \cmark & \cmark & \cmark& \cmark & \NA \\  \hline
	$\mathbf{cosh(x)}$ & \cmark & \cmark & \cmark & \cmark & \NA  \\  \hline
	$\mathbf{sinpi(x)}$ & \cmark & \cmark& \cmark & \cmark & \NA  \\ \hline
	$\mathbf{cospi(x)}$ & \cmark & \cmark & \cmark & \cmark & \NA  \\ \hline
\end{tabular}
\begin{tabular}{| P{1.3em} | P{1.3em} | P{1.3em} | P{1.3em} | P{1.3em} |}
	\hline
	\multicolumn{5}{| c |}{Using \Rlibmtt} \\
	\hline
	\hline
	\RNE & \RNN & \RNP & \RNZ & \RNA\\
	\hline
	\xmark & \xmark & \xmark & \xmark & \xmark \\ \hline
	\cmark & \cmark & \cmark & \cmark & \cmark  \\ \hline
	\xmark & \xmark & \xmark & \xmark & \xmark  \\ \hline
	\xmark & \xmark & \xmark & \xmark & \xmark \\ \hline
	\xmark & \xmark & \xmark & \xmark & \cmark  \\ \hline
	\cmark & \xmark & \xmark& \xmark & \xmark  \\ \hline
	\xmark & \xmark & \xmark& \xmark & \xmark \\  \hline
	\xmark & \xmark & \xmark & \xmark & \xmark  \\  \hline
	\cmark & \xmark& \xmark & \xmark & \cmark  \\ \hline
	\cmark & \xmark & \xmark & \xmark & \cmark  \\ \hline
\end{tabular}

\label{tbl:eval:ofa:tf32acc}
\end{table*}

%% file: tbl.eval.ofa.bf16acc.tex
\begin{table*}
	\footnotesize
	\caption{\small Generation of correctly rounded results for
          bfloat16 using the five standard IEEE-754 rounding modes
          \RNE, \RNN, \RNP, \RNZ, and \RNA. We show the results with
          the elementary functions from \tool, glibc's double libm,
          Intel's double libm, \Rlibmtt, and CR-LIBM. Then, we convert
          the output to bfloat16 values. \cmark indicates that the
          library produces the correctly rounded bfloat16 result using
          a given rounding mode for all inputs. Otherwise, we use
          \xmark. We use the separate approximation provided for each
          of the four rounding modes with CR-LIBM.}
\begin{tabular}{| P{3.5em} | P{1.3em} | P{1.3em} | P{1.3em} | P{1.3em} | P{1.3em} |}
	\hline
	& \multicolumn{5}{| c |}{\textbf{Using \tool}} \\
	\hline
	\hline
	$f(x)$ & \RNE & \RNN & \RNP & \RNZ & \RNA\\
	\hline
	$\mathbf{ln(x)}$ & \cmark & \cmark & \cmark & \cmark & \cmark \\ \hline
	$\mathbf{log_2(x)}$ & \cmark & \cmark & \cmark & \cmark & \cmark  \\ \hline
	$\mathbf{log_{10}(x)}$ & \cmark & \cmark & \cmark & \cmark & \cmark  \\ \hline
	$\mathbf{e^{x}}$ & \cmark & \cmark & \cmark & \cmark & \cmark \\ \hline
	$\mathbf{2^{x}}$ & \cmark & \cmark & \cmark & \cmark & \cmark  \\ \hline
	$\mathbf{10^{x}}$ & \cmark & \cmark & \cmark& \cmark & \cmark  \\ \hline
	$\mathbf{sinh(x)}$ & \cmark & \cmark & \cmark& \cmark & \cmark \\  \hline
	$\mathbf{cosh(x)}$ & \cmark & \cmark & \cmark & \cmark & \cmark  \\  \hline
	$\mathbf{sinpi(x)}$ & \cmark & \cmark& \cmark & \cmark & \cmark  \\ \hline
	$\mathbf{cospi(x)}$ & \cmark & \cmark & \cmark & \cmark & \cmark  \\ \hline
\end{tabular}
\begin{tabular}{| P{1.3em} | P{1.3em} | P{1.3em} | P{1.3em} | P{1.3em} |}
	\hline
	\multicolumn{5}{| c |}{Using glibc double libm} \\
	\hline
	\hline
	\RNE & \RNN & \RNP & \RNZ & \RNA\\
	\hline
	\cmark & \cmark & \cmark & \cmark & \cmark \\ \hline
	\cmark & \cmark & \cmark & \cmark & \cmark  \\ \hline
	\cmark & \cmark & \cmark & \cmark & \cmark  \\ \hline
	\cmark & \xmark & \xmark & \xmark & \cmark \\ \hline
	\cmark & \xmark & \xmark& \xmark & \cmark  \\ \hline
	\cmark & \xmark & \xmark & \xmark & \cmark  \\ \hline
	\cmark & \xmark & \xmark& \xmark & \cmark \\  \hline
	\cmark & \xmark & \xmark & \xmark & \cmark  \\  \hline
	\NA & \NA& \NA & \NA & \NA  \\ \hline
	\NA & \NA & \NA & \NA & \NA  \\ \hline
\end{tabular}
\begin{tabular}{| P{1.3em} | P{1.3em} | P{1.3em} | P{1.3em} | P{1.3em} |}
	\hline
	\multicolumn{5}{| c |}{Using Intel double libm} \\
	\hline
	\hline
	\RNE & \RNN & \RNP & \RNZ & \RNA\\
	\hline
	\cmark & \cmark & \cmark & \cmark & \cmark \\ \hline
	\cmark & \cmark & \cmark & \cmark & \cmark  \\ \hline
	\cmark & \cmark & \cmark & \cmark & \cmark  \\ \hline
	\cmark & \xmark & \xmark & \xmark & \cmark \\ \hline
	\cmark & \xmark & \xmark & \xmark & \cmark  \\ \hline
	\cmark & \xmark & \xmark& \xmark & \cmark  \\ \hline
	\cmark & \xmark & \xmark& \xmark & \cmark \\  \hline
	\cmark & \xmark & \xmark & \xmark & \cmark  \\  \hline
	\cmark & \cmark& \cmark & \cmark & \cmark  \\ \hline
	\cmark & \xmark & \cmark & \xmark & \cmark  \\ \hline
\end{tabular}
\begin{tabular}{| P{1.3em} | P{1.3em} | P{1.3em} | P{1.3em} | P{1.3em} |}
	\hline
	\multicolumn{5}{| c |}{Using CRLIBM} \\
	\hline
	\hline
	\RNE & \RNN & \RNP & \RNZ & \RNA\\
	\hline
	\cmark & \cmark & \cmark & \cmark & \NA \\ \hline
	\cmark & \cmark & \cmark & \cmark & \NA  \\ \hline
	\cmark & \cmark & \cmark & \cmark & \NA  \\ \hline
	\cmark & \cmark & \cmark & \cmark & \NA \\ \hline
	\NA & \NA& \NA & \NA & \NA  \\ \hline
	\NA& \NA & \NA & \NA & \NA  \\ \hline
	\cmark & \cmark & \cmark& \cmark & \NA \\  \hline
	\cmark & \cmark & \cmark & \cmark & \NA  \\  \hline
	\cmark & \cmark& \cmark & \cmark & \NA  \\ \hline
	\cmark & \cmark & \cmark & \cmark & \NA \\ \hline
\end{tabular}
\begin{tabular}{| P{3.5em} | P{1.3em} | P{1.3em} | P{1.3em} | P{1.3em} | P{1.3em} |}
	\hline
	& \multicolumn{5}{| c |}{Using \Rlibmtt} \\
	\hline
	\hline
	$f(x)$ & \RNE & \RNN & \RNP & \RNZ & \RNA\\
	\hline
	$\mathbf{ln(x)}$ & \cmark & \xmark & \xmark & \cmark & \xmark \\ \hline
	$\mathbf{log_2(x)}$ & \cmark & \cmark & \cmark & \cmark & \cmark  \\ \hline
	$\mathbf{log_{10}(x)}$ & \cmark & \cmark & \cmark & \cmark & \cmark  \\ \hline
	$\mathbf{2^x}$ & \cmark & \xmark & \xmark & \xmark & \cmark \\ \hline
	$\mathbf{2^x}$ & \cmark & \xmark & \xmark & \xmark & \cmark  \\ \hline
	$\mathbf{10^x}$ & \xmark & \xmark & \xmark& \xmark & \xmark  \\ \hline
	$\mathbf{sinh(x)}$ & \cmark & \xmark & \xmark& \xmark & \cmark \\  \hline
	$\mathbf{cosh(x)}$ & \cmark & \xmark & \xmark & \xmark & \cmark  \\  \hline
	$\mathbf{sinpi(x)}$ & \cmark & \cmark& \cmark & \cmark & \cmark  \\ \hline
	$\mathbf{cospi(x)}$ & \cmark & \xmark & \cmark & \xmark & \cmark  \\ \hline
\end{tabular}
\label{tbl:eval:ofa:bf16acc}
\end{table*}

%% file: fig.eval.ofa.GrlibmFloatSpeedup.tex
\begin{figure*}
  \small
  \begin{subfigure}[b]{0.49\columnwidth}
    \caption{Speedup of \tool over glibc's libm}
    \includegraphics[width=\linewidth]{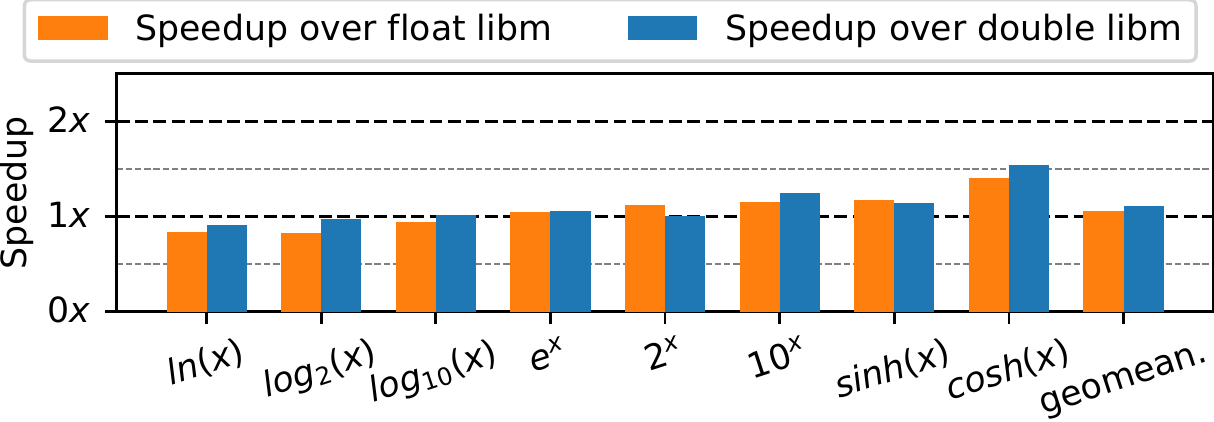}
  \end{subfigure}
    \hfill
  \begin{subfigure}[b]{0.49\columnwidth}
    \caption{Speedup of \tool over Intel's libm}
    \includegraphics[width=\linewidth]{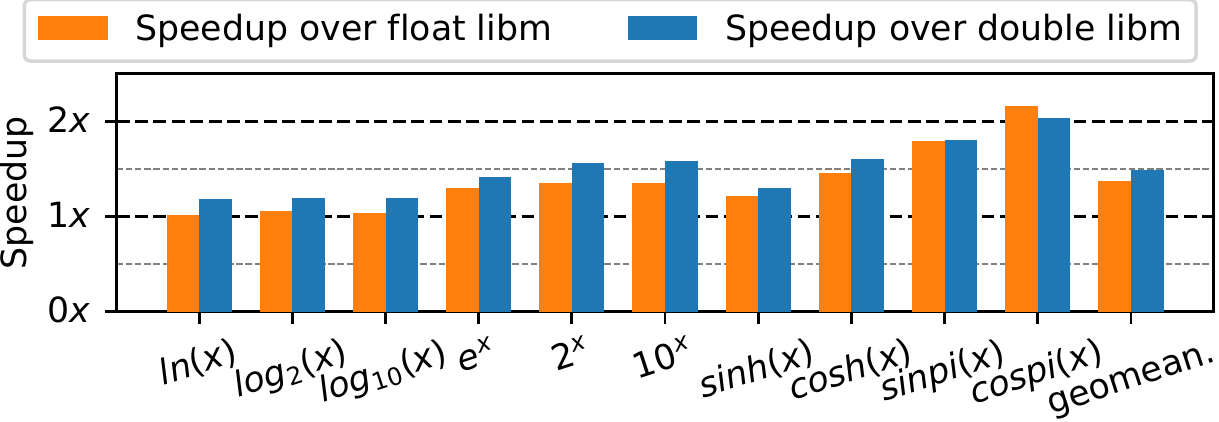}
  \end{subfigure}
  \vspace{1em}  
  \begin{subfigure}[b]{0.49\columnwidth}
    \caption{Speedup of \tool over CR-LIBM}
    \includegraphics[width=\linewidth]{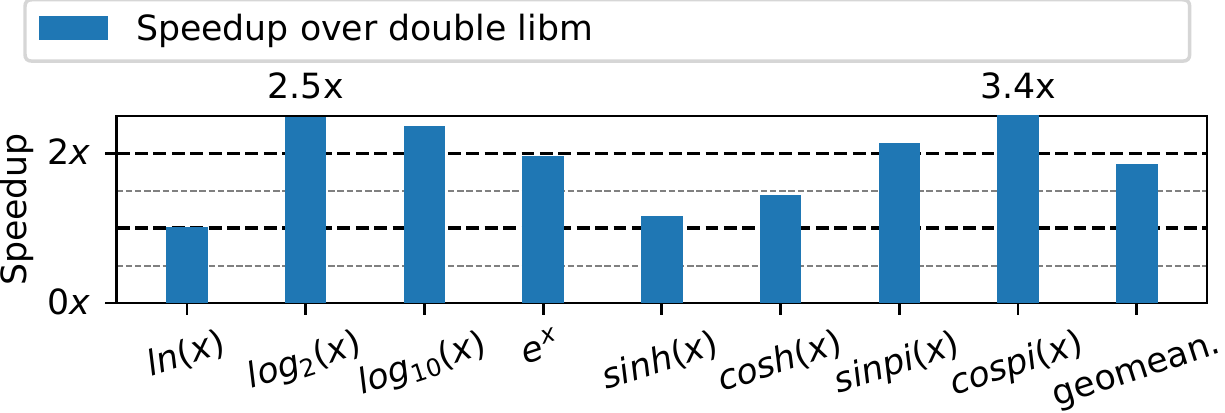}
  \end{subfigure}
  \hfill  
  \begin{subfigure}[b]{0.49\columnwidth}
    \caption{Speedup of \tool over \Rlibmtt}
    \includegraphics[width=\linewidth]{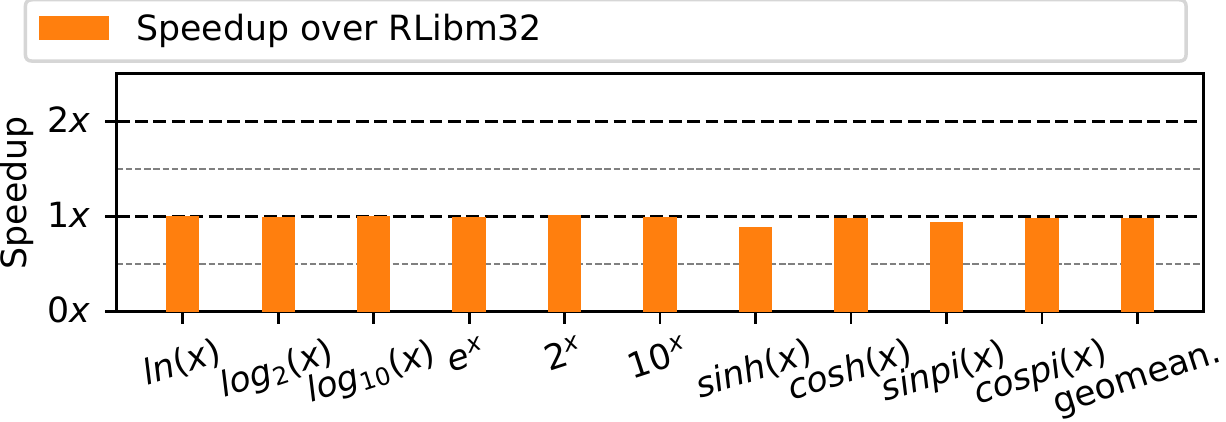}
  \end{subfigure}
  \caption{\small (a) Speedup of \tool's functions compared to glibc's
    float functions (left) and glibc's double functions (right) when
    producing 32-bit float results.  (b) Speedup of \tool's functions
    compared to Intel's float functions (left) and Intel's double
    functions (right) when producing 32-bit float results. (c) Speedup
    of \tool's functions compared to CR-LIBM functions when producing
    32-bit float results. (d) Speedup of \tool's functions compared to
    \Rlibmtt functions when producing 32-bit float results.  }
  \label{fig:eval:ofa:GrlibmFloatSpeedup}
\end{figure*}

%% file: sec.related.tex
\section{Related Work}
Seminal research over multiple decades has advanced the
state-of-the-art for creating polynomial
approximations~\cite{Jeannerod:sqrt:tc:2011, Bui:exp:ccece:1999,
  Abraham:fastcorrect:toms:1991, 
  Daramy:crlibm:spie:2003,Fousse:toms:2007:mpfr,Muller:elemfunc:book:2005,Trefethen:chebyshev:book:2012,
  Remes:algorithm:1934}. Important advances in range reduction has
made such approximation
feasible~\cite{Tang:log:toms:1990,Tang:TableLookup:SCA:1991,Tang:exp:toms:1989,762822,
  Cody:book:1980, Boldo:reduction:toc:2009}.
Simultaneously, there are verification efforts to prove bounds for
math libraries~\cite{harrison:hollight:tphols:2009,
  Harrison:expproof:amst:1997, Harrison:verifywithHOL:tphol:1997,
  Sawada:verify:acl:2002,Lee:verify:popl:2018} and repair individual
outputs of math
libraries~\cite{Xin:repairmlib:popl:2019,Daming:fpe:popl:2020}.
%
%
The comprehensive book on elementary functions provides detailed
information on prior work~\cite{Muller:elemfunc:book:2005}.

We restrict our comparison to prior work that is closely related to
our work.  As a correctly rounded elementary function is recommended
by the IEEE standard and enables portability, a number of correctly
rounded math libraries have been
developed~\cite{Abraham:fastcorrect:toms:1991,
  Daramy:crlibm:spie:2003, lim:rlibm:popl:2021,Lim:rlibm:arxiv:2020,
  lim:rlibm32:pldi:2021}. They are restricted to a specific
representation and a rounding mode.

CR-LIBM~\cite{Daramy:crlibm:spie:2003, Lefevre:toward:tc:1998} is a
correctly rounded collection of elementary functions for double
precision. It was developed using
Sollya~\cite{Chevillard:sollya:icms:2010}, which is a tool and a
library for developing FP code. Sollya can generate polynomials of
degree $d$ with coefficients in a representation used for the
implementation ($\mathbb{H}$) that has the minimum infinity
norm~\cite{Brisebarre:epl:arith:2007}. Sollya uses a modified Remez
algorithm to produce polynomials. It also computes and proves the
error bound on the polynomial evaluation using interval
arithmetic~\cite{Chevillard:infnorm:qsic:2007,
  Chevillard:ub:tcs:2011}.  Metalibm~\cite{Olga:metalibm:icms:2014,
  Brunie:metalibm:ca:2015} is a customization infrastructure also
built using Sollya. MetaLibm is able to automatically identify range
reduction and domain splitting techniques for some transcendental
functions. It has been used to trade-off correctness and performance
while approximating elementary functions for float and double
precision types.

A modified Remez algorithm has also been used to generate polynomials
that minimizes the infinity norm compared to an ideal elementary
function~\cite{Arzelier:poly:arith:2019}. It can be useful for
generating correctly rounded results for a specific precision and a
rounding mode when range reduction is not necessary.

This paper builds on our prior work in the \rlibm
project~\cite{lim:rlibm:popl:2021,Lim:rlibm:arxiv:2020,
  lim:rlibm32:pldi:2021, Lim:rlibmAll:arxiv:2021,
  lim:rlibm:phdthesis:2021} that creates polynomials using the
correctly rounded value rather than the real value of the elementary
function.  Like the \rlibm project, we structure the problem of
generating polynomials as an LP problem. We also use \rlibm's range
reduction strategies.  The \rlibm project has generated correctly
rounded libraries with the commonly used \RNE mode for multiple types:
bfloat16, posit16, 32-bit float, and posit32. However, it is necessary
to create individual polynomial approximation for each representation
with each rounding mode with the \rlibm project to avoid double
rounding errors.  In contrast, this paper shows that by generating
polynomial approximations for \Tlarge with the round-to-odd mode, we
can create a single polynomial approximation that works for multiple
representations \Tsmall with multiple rounding modes.

%% file: sec.conclusion.tex
\section{Conclusion}
This paper proposes a novel method to generate a single polynomial
approximation that produces correctly rounded results for multiple
representations and rounding modes. The key idea is to create a
polynomial approximation that produces the correctly rounded result
for \Tlarge with the round-to-odd mode when the goal is to generate
correct results for \Tsmall, where $k \leq n$, with all rounding
modes. We address the issue of singletons while generating polynomials
that approximate the correctly rounded result with the round-to-odd
mode. We provide the first correctly rounded implementations of
elementary functions for multiple representations.  We believe that
our results make a strong case for mandating correctly rounded results
at least with any representation that has fewer than or equal to
32-bits.